\documentclass{article}

\makeatletter

\usepackage[verbose=true,letterpaper]{geometry}
\AtBeginDocument{
  \newgeometry{
    textheight=9in,
    textwidth=6.5in,
    top=1in,
    headheight=14pt,
    headsep=25pt,
    footskip=30pt
  }
}

\renewcommand{\normalsize}{%
  \@setfontsize\normalsize\@xpt\@xipt
  \abovedisplayskip      7\p@ \@plus 2\p@ \@minus 5\p@
  \abovedisplayshortskip \z@ \@plus 3\p@
  \belowdisplayskip      \abovedisplayskip
  \belowdisplayshortskip 4\p@ \@plus 3\p@ \@minus 3\p@
}
\normalsize
\renewcommand{\small}{%
  \@setfontsize\small\@ixpt\@xpt
  \abovedisplayskip      6\p@ \@plus 1.5\p@ \@minus 4\p@
  \abovedisplayshortskip \z@  \@plus 2\p@
  \belowdisplayskip      \abovedisplayskip
  \belowdisplayshortskip 3\p@ \@plus 2\p@   \@minus 2\p@
}
\renewcommand{\footnotesize}{\@setfontsize\footnotesize\@ixpt\@xpt}
\renewcommand{\scriptsize}{\@setfontsize\scriptsize\@viipt\@viiipt}
\renewcommand{\tiny}{\@setfontsize\tiny\@vipt\@viipt}
\renewcommand{\large}{\@setfontsize\large\@xiipt{14}}
\renewcommand{\Large}{\@setfontsize\Large\@xivpt{16}}
\renewcommand{\LARGE}{\@setfontsize\LARGE\@xviipt{20}}
\renewcommand{\huge}{\@setfontsize\huge\@xxpt{23}}
\renewcommand{\Huge}{\@setfontsize\Huge\@xxvpt{28}}

\providecommand{\section}{}
\renewcommand{\section}{%
  \@startsection{section}{1}{\z@}%
                {-2.0ex \@plus -0.5ex \@minus -0.2ex}%
                { 1.5ex \@plus  0.3ex \@minus  0.2ex}%
                {\large\bf\raggedright}%
}
\providecommand{\subsection}{}
\renewcommand{\subsection}{%
  \@startsection{subsection}{2}{\z@}%
                {-1.8ex \@plus -0.5ex \@minus -0.2ex}%
                { 0.8ex \@plus  0.2ex}%
                {\normalsize\bf\raggedright}%
}
\providecommand{\subsubsection}{}
\renewcommand{\subsubsection}{%
  \@startsection{subsubsection}{3}{\z@}%
                {-1.5ex \@plus -0.5ex \@minus -0.2ex}%
                { 0.5ex \@plus  0.2ex}%
                {\normalsize\bf\raggedright}%
}
\providecommand{\paragraph}{}
\renewcommand{\paragraph}{%
  \@startsection{paragraph}{4}{\z@}%
                {1.5ex \@plus 0.5ex \@minus 0.2ex}%
                {-1em}%
                {\normalsize\bf}%
}
\providecommand{\subparagraph}{}
\renewcommand{\subparagraph}{%
  \@startsection{subparagraph}{5}{\z@}%
                {1.5ex \@plus 0.5ex \@minus 0.2ex}%
                {-1em}%
                {\normalsize\bf}%
}

\newlength{\@abovecaptionskip}
\newlength{\@belowcaptionskip}

\renewenvironment{table}
  {\setlength{\abovecaptionskip}{\@belowcaptionskip}%
   \setlength{\belowcaptionskip}{\@abovecaptionskip}%
   \@float{table}}
  {\end@float}

\renewcommand{\footnoterule}{\kern-3\p@ \hrule width 12pc \kern 2.6\p@}
\def\@listi  {\leftmargin\leftmargini}
\def\@listii {\leftmargin\leftmarginii
              \labelwidth\leftmarginii
              \advance\labelwidth-\labelsep
              \topsep  2\p@ \@plus 1\p@    \@minus 0.5\p@
              \parsep  1\p@ \@plus 0.5\p@ \@minus 0.5\p@
              \itemsep \parsep}
\def\@listiii{\leftmargin\leftmarginiii
              \labelwidth\leftmarginiii
              \advance\labelwidth-\labelsep
              \topsep    1\p@ \@plus 0.5\p@ \@minus 0.5\p@
              \parsep    \z@
              \partopsep 0.5\p@ \@plus 0\p@ \@minus 0.5\p@
              \itemsep \topsep}
\def\@listiv {\leftmargin\leftmarginiv
              \labelwidth\leftmarginiv
              \advance\labelwidth-\labelsep}
\def\@listv  {\leftmargin\leftmarginv
              \labelwidth\leftmarginv
              \advance\labelwidth-\labelsep}
\def\@listvi {\leftmargin\leftmarginvi
              \labelwidth\leftmarginvi
              \advance\labelwidth-\labelsep}

\providecommand{\maketitle}{}
\renewcommand{\maketitle}{%
  \par
  \begingroup
    \renewcommand{\thefootnote}{\fnsymbol{footnote}}
    \renewcommand{\@makefnmark}{\hbox to \z@{$^{\@thefnmark}$\hss}}
    \long\def\@makefntext##1{%
      \parindent 1em\noindent
      \hbox to 1.8em{\hss $\m@th ^{\@thefnmark}$}##1
    }
    \thispagestyle{empty}
    \@maketitle
    \@thanks
  \endgroup
  \let\maketitle\relax
  \let\thanks\relax
}

\newcommand{\@toptitlebar}{
  \hrule height 2\p@
  \vskip 0.25in
  \vskip -\parskip%
}
\newcommand{\@bottomtitlebar}{
  \vskip 0.29in
  \vskip -\parskip
  \hrule height 2\p@
  \vskip 0.09in%
}

\providecommand{\@maketitle}{}
\renewcommand{\@maketitle}{%
  \vbox{%
    \hsize\textwidth
    \linewidth\hsize
    \vskip 0.1in
    \@toptitlebar
    \centering
    {\LARGE\sc \@title\par}
    \@bottomtitlebar
    \vskip 0.1in
    \def\And{%
      \end{tabular}\hfil\linebreak[0]\hfil%
      \begin{tabular}[t]{c}\bf\rule{\z@}{24\p@}\ignorespaces%
    }
    \def\AND{%
      \end{tabular}\hfil\linebreak[4]\hfil%
      \begin{tabular}[t]{c}\bf\rule{\z@}{24\p@}\ignorespaces%
    }
    \begin{tabular}[t]{c}\bf\rule{\z@}{24\p@}\@author\end{tabular}%
  \vskip 0.4in \@minus 0.1in \center{ }   \vskip 0.2in 
  }
}

\renewenvironment{abstract}
{
  \centerline
  {\large \bfseries \scshape Abstract}
  \begin{quote}
}
{
  \end{quote}
}

\makeatother

\usepackage[utf8]{inputenc}
\usepackage[T1]{fontenc}
\usepackage{textcomp}
\usepackage{hyperref}
\usepackage{url}
\usepackage{booktabs}
\usepackage{amsfonts}
\usepackage{nicefrac}
\usepackage{microtype}
\usepackage{lipsum}
\usepackage{float}
\usepackage{natbib}
\usepackage{appendix}
\usepackage{fancyhdr}
\usepackage{graphicx}
\graphicspath{{media/}}
\usepackage{wrapfig}
\usepackage[most]{tcolorbox}
\usepackage{amsmath}
\usepackage{amssymb}
\usepackage{bbm}
\usepackage{longtable}
\usepackage{array}
\usepackage{multirow}
\usepackage{subcaption}
\usepackage{multicol}
\usepackage{pdfpages}
\usepackage{cleveref}

\usepackage{listings}
\definecolor{lstbg}{RGB}{247,248,250}
\definecolor{lstkw}{RGB}{0,86,148}
\definecolor{lstcm}{RGB}{112,124,136}
\definecolor{lststr}{RGB}{168,60,42}
\definecolor{lstrule}{RGB}{203,209,216}
\lstdefinestyle{agentpy}{
  language=Python,
  basicstyle=\ttfamily\scriptsize,
  keywordstyle=\color{lstkw}\bfseries,
  commentstyle=\color{lstcm}\itshape,
  stringstyle=\color{lststr},
  backgroundcolor=\color{lstbg},
  frame=single, rulecolor=\color{lstrule}, framesep=3pt,
  xleftmargin=16pt, xrightmargin=1pt,
  numbers=left, numberstyle=\ttfamily\tiny\color{lstcm}, numbersep=7pt,
  breaklines=true, breakatwhitespace=false,
  postbreak=\mbox{\textcolor{lstcm}{$\hookrightarrow$}\space},
  showstringspaces=false, columns=fullflexible, keepspaces=true,
  aboveskip=3pt, belowskip=1pt,
  extendedchars=true,
  literate={×}{{\texttimes}}1 {≤}{{$\leq$}}1 {≥}{{$\geq$}}1 {∪}{{$\cup$}}1 {—}{{---}}1 {→}{{$\rightarrow$}}1,
}
\lstdefinestyle{agentout}{
  basicstyle=\ttfamily\scriptsize,
  backgroundcolor=\color{white},
  frame=single, rulecolor=\color{lstrule}, framesep=3pt,
  xleftmargin=2pt, xrightmargin=1pt,
  breaklines=true, breakatwhitespace=false,
  postbreak=\mbox{\textcolor{lstrule}{$\hookrightarrow$}\space},
  showstringspaces=false, columns=fullflexible, keepspaces=true,
  aboveskip=3pt, belowskip=1pt,
}

\usepackage{amssymb}
\usepackage{authblk}   
\usepackage{xcolor}
\usepackage{xspace}    

\newcommand{\needcite}[1][]{\textcolor{red}{[\textbf{CITE\ifx&#1&\else: #1\fi}]}}

\newcommand{\method}{\textsc{AI Sleep Co-Scientist}\xspace}

\newcommand{\needstat}[1][]{\textcolor{red}{[\textbf{STAT\ifx&#1&\else: #1\fi}]}}

\definecolor{mfblue}{RGB}{49,130,189}
\newtcolorbox{mainfinding}{
  enhanced,
  colback=mfblue!8,
  colframe=mfblue!85!black,
  colbacktitle=mfblue!22,
  coltitle=black,
  fonttitle=\bfseries\small,
  title=Main finding,
  boxrule=0.5pt,
  arc=1pt,
  left=8pt, right=8pt, top=4pt, bottom=4pt,
  before skip=10pt, after skip=10pt,
}

\definecolor{mfnote}{RGB}{200,90,30}

\newtcolorbox{expertnote}{
  enhanced,
  colback=mfnote!6,
  colframe=mfnote!85!black,
  colbacktitle=mfnote!22,
  coltitle=black,
  fonttitle=\bfseries\small,
  title=Note to expert collaborators (remove before submission),
  boxrule=0.5pt,
  arc=1pt,
  left=8pt, right=8pt, top=4pt, bottom=4pt,
  before skip=10pt, after skip=10pt,
}

\newtcolorbox{collaboration}{
  enhanced,
  colback=gray!6,
  colframe=gray!55!black,
  colbacktitle=gray!22,
  coltitle=black,
  fonttitle=\bfseries\small,
  title=Expert note on collaboration,
  boxrule=0.5pt,
  arc=1pt,
  left=8pt, right=8pt, top=4pt, bottom=4pt,
  before skip=10pt, after skip=10pt,
}

\newcounter{extfigure}
\newcounter{exttable}

\title{Agentic AI-enabled discovery across large-scale sleep physiology}

\author[1,2]{Rahul Thapa}
\author[3,4,*]{Umaer Hanif}
\author[5,*]{Robin Guillard}
\author[3,*]{Andreas Brink-Kjaer}
\author[5,6,*]{Adrien Specht}
\author[3,*]{Matteo Saibene}
\author[3,4,5]{Magnus Ruud Kjaer}
\author[1]{Harrison G. Zhang}
\author[7]{Federico Bianchi}
\author[5]{Elisabeth Roxane M. Heremans}
\author[8,9]{Eric C. Landsness}
\author[5,+]{Emmanuel Mignot}
\author[1,2,+]{James Zou}

\affil[1]{Department of Biomedical Data Science, Stanford University, Stanford, CA, USA}
\affil[2]{Department of Computer Science, Stanford University, Stanford, CA, USA}
\affil[3]{Department of Health Technology, Technical University of Denmark, Kongens Lyngby, Denmark}
\affil[4]{Danish Center for Sleep Medicine, Department of Clinical Neurophysiology, Glostrup, Denmark}
\affil[5]{Department of Psychiatry and Behavioral Sciences, Stanford University, Stanford, CA, USA}
\affil[6]{Institute for Computational and Mathematical Engineering, Stanford University, Stanford, CA, USA}
\affil[7]{Together AI, USA}
\affil[8]{Department of Neurology, Washington University School of Medicine in St. Louis, MO, USA}
\affil[9]{Consortium for Biomedical Research and AI in Neurodegeneration (C-BRAIN), USA}

\affil[*]{Each led one case study and contributed equally. \quad \textsuperscript{+}Jointly supervised this work.}
\affil[ ]{\textbf{Correspondence:} \href{mailto:rthapa84@stanford.edu}{rthapa84@stanford.edu}, \href{mailto:mignot@stanford.edu}{mignot@stanford.edu}, \href{mailto:jamesz@stanford.edu}{jamesz@stanford.edu}}

\begin{document}

\maketitle

\begin{abstract}
Sleep occupies roughly one-third of human life and plays a central role in human health, yet many aspects of its physiology remain poorly understood. Large polysomnography (PSG) datasets offer new opportunities to study disease risk, clinical phenotypes, and the biological organization of sleep, but extracting insight from these complex, multimodal recordings requires substantial expert effort and remains difficult for general-purpose AI systems. We therefore developed \method, an expert-guided research environment in which human scientists direct specialist agents for hypothesis development, signal preprocessing, and statistical analysis, while reviewing intermediate outputs. Each reported result is linked to the executable code that produced it, providing an auditable record of the research process. Across four clinical and epidemiological cohorts comprising approximately 124{,}000 PSG recordings and more than 50\,TB of raw signals, we conducted five case studies spanning three directions: how sleep physiology relates to future disease, how it distinguishes clinical phenotypes, and how sleep is organized and regulated. Diminished network-level physiological coupling during sleep was associated with incident Parkinson's disease (HR 1.48, 95\% CI 1.31--1.67) and Alzheimer's disease (HR 1.38, 95\% CI 1.25--1.53). A physiologically structured late-fusion sleep-age model outperformed an unconstrained early-fusion approach (mean absolute error 7.06 vs.\ 7.33 years in the Stanford Sleep Clinic cohort and 8.83 vs.\ 9.52 years in the Human Sleep Project), and its age residual was associated with incident disease across multiple organ systems. Arousal dynamics characterized comorbid insomnia and sleep apnoea as an intermediate phenotype skewed towards obstructive sleep apnoea, but distinguished from it by prolonged post-arousal wakefulness and more irregular arousal organization. Analyses of sleep regulation showed that rapid eye movement (REM) bout duration was more strongly associated with the amount of non-REM sleep than with intervening wakefulness. Transitions from non-REM to REM sleep were least likely when the probability of entering deep sleep peaked, in both narcolepsy type~1 and controls. Finally, transient-oscillation analysis identified a fast-sigma deficit and excess centrofrontal theta activity in narcolepsy type~1. Together, these findings connect features of sleep to disease risk, clinical classification, and the regulation of sleep itself, and demonstrate how agentic AI can support large-scale, multimodal discovery.
\end{abstract}

\section*{Introduction}

Humans spend roughly one-third of their lives asleep, and disordered or insufficient sleep is consistently associated with elevated risk of cardiovascular disease, type~2 diabetes, depression, and all-cause mortality~\citep{cappuccio2010sleep, lim2010meta}.
Yet the biological functions of sleep remain only partially understood~\citep{cirelli2008sleep, rechtschaffen1983physiological, walker2017we}.
The clinical gold standard, polysomnography (PSG), captures a full night of physiology as a rich, multichannel recording~\citep{berry2012aasm, nasiri2025caisr}, and decades of clinical practice have accumulated vast PSG archives.
Manual scoring and the high dimensionality of these recordings have nonetheless confined most studies to a handful of pre-specified hypotheses, leaving much of what these archives hold unexamined.

Sleep foundation models and related deep-learning tools have shown that a single night of PSG carries signals predictive of clinical outcomes years later, establishing that these archives are worth mining at scale~\citep{thapa2024sleepfm, thapa2026multimodal, kjaer2025stanford, xu2026sleeplm}. These models, though, are built to improve clinical prediction rather than to explain the physiology behind it, leaving a different class of question open. Which features within the sleep signal correlate best with which outcomes? What drives those correlations, and is it possible to move from correlation towards mechanism? Where conventional clinical categories are contested, can physiology adjudicate between them? Each of these questions requires analysing large, multimodal cohorts end to end, and it is the cost of doing so that has left the archives underused.

That cost has three sources. The first is scale: a single PSG contains hours of continuously sampled, multichannel physiology, and modern clinical archives can hold tens of thousands of such recordings~\citep{kjaer2025stanford}, so a single hypothesis may require processing hundreds of thousands of hours of signal. The second is multimodal complexity: each recording integrates concurrent channels spanning the brain, autonomic, respiratory, and musculoskeletal systems, sampled at up to hundreds of hertz, each requiring distinct clinical interpretation. The third is fragmentation: recordings accumulate at different institutions under different montages and channel-naming conventions, and must be harmonized before they can be analysed together. Layered over all three is the domain expertise the data demand, spanning consensus scoring standards, sleep physiology, and the epidemiological methods needed to link one night of PSG to a diagnosis recorded years later.

Research agents built on large language models have begun to support several stages of this kind of work, including literature synthesis, hypothesis generation, code execution, data analysis, and reporting, with the shared aim of widening the range of hypotheses a scientist can examine while reducing the time spent on implementation and iterative testing~\citep{boiko2023autonomous,lu2026towards,mitchener2025kosmos,huang2026autonomous,tang2026ai,lu2026eubiota,zhang2026virtual,novikov2025alphaevolve,gottweis2026accelerating,gao2024empowering}. Some specialize in long-horizon autonomous execution from a researcher-supplied scientific objective~\citep{lu2026towards,mitchener2025kosmos,ifargan2025autonomous}; others emphasize ideation and multi-agent collaboration under human direction~\citep{gottweis2026accelerating,swanson2025virtual,zhang2026virtual}; others connect model reasoning to specialized scientific software and databases~\citep{boiko2023autonomous,huang2026autonomous,lu2026eubiota,novikov2025alphaevolve}. These demonstrations have largely operated on literature-grounded hypothesis spaces, standard machine-learning datasets, or preprocessed tabular and matrix data rather than cohort-scale archives of raw physiological signal~\citep{lu2026towards,gottweis2026accelerating,mitchener2025kosmos,huang2026autonomous,ifargan2025autonomous,swanson2025virtual}.

Two existing agentic systems illustrate the gap. Biomni gives an agent access to a large library of biomedical tools, software packages, and curated databases. That design works because much of molecular biology is already standardized: single-cell analysis has agreed data formats, ready-made pipelines an agent can install and run, and reference databases that define the genes under study. Sleep physiology has not converged to the same degree. Its measurements must first be derived from raw multichannel recordings using validated detectors for sleep staging, arousals, respiratory events, and oscillatory microstructure, applied to signals whose channel names and montages differ from site to site. Deriving them is part of the scientific work rather than a step that precedes it, and Biomni offers no pipelines for overnight polysomnography, drawing on curated flat files rather than raw signal~\citep{huang2026autonomous}. Kosmos sustains long autonomous cycles of analysis and literature search, but over supplied datasets of a few gigabytes rather than raw signal, and it fixes its research direction at the outset, leaving researchers little opportunity to intervene between cycles to steer it~\citep{mitchener2025kosmos}. The archives analysed here are orders of magnitude larger, and every feature used in these case studies had to be derived from raw signal. Where scoring conventions, preprocessing, channel availability, and clinical plausibility all require expert judgement partway through an analysis, limited steerability constrains alignment between system and investigator as much as scale does. Closing this gap depends less on a more capable general-purpose agent than on the infrastructure built around one.

\begin{figure}[!htbp]
    \centering
    \includegraphics[width=\linewidth]{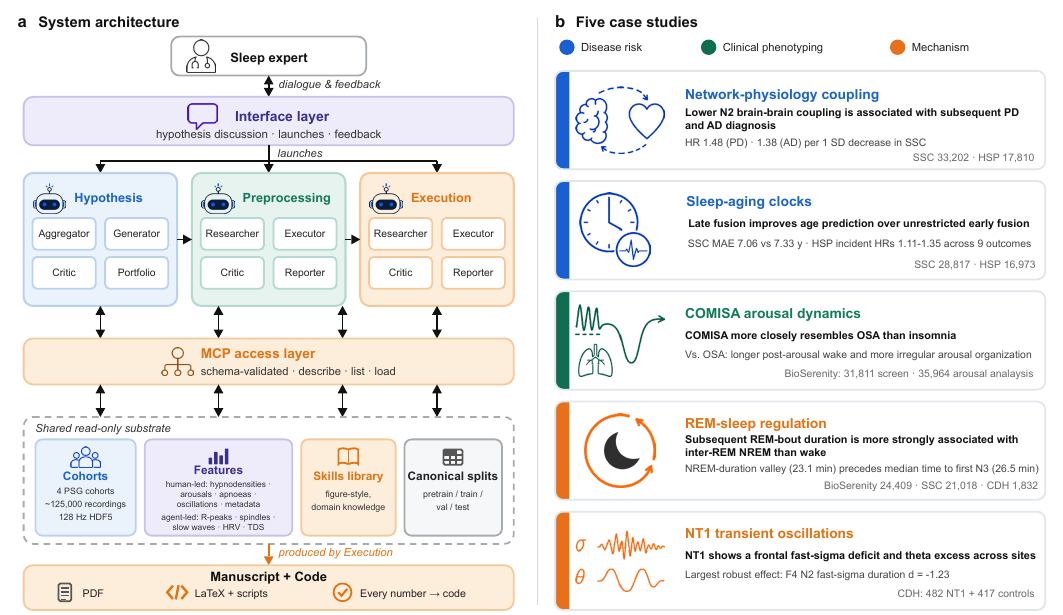}
    \caption{\textbf{Overview of \method.}
    \textbf{(a)} System architecture. A sleep expert directs research through an \emph{Interface Layer}, a conversational application for hypothesis discussion, agent launches, and structured feedback. The interface coordinates three specialist agents: a \emph{Hypothesis Agent} that generates, critiques, and aggregates candidate hypotheses into a portfolio; a \emph{Preprocessing Agent} that develops and validates physiological feature pipelines; and an \emph{Execution Agent} that carries selected hypotheses through analysis to a code-grounded manuscript. The Hypothesis Agent comprises Aggregator, Generator, Critic, and Portfolio sub-agents, while the Preprocessing and Execution Agents each comprise Researcher, Executor, Critic, and Reporter sub-agents. The specialist agents access a shared read-only substrate through a schema-validated Model Context Protocol (MCP) layer supporting data description, listing, and loading. The substrate contains four PSG cohorts ($\sim$124{,}000 recordings; $>50$\,TB), curated physiological features, a shared skills library, and canonical train/validation/test splits. The Execution Agent returns a manuscript and the associated \LaTeX  \ files, analysis scripts, and executable provenance.
    \textbf{(b)} Five case studies spanning disease risk, clinical phenotyping, and mechanism, each summarized by its principal finding, supporting result, and analysis cohorts.}
    \label{fig:overview}
\end{figure}

We therefore built \method, designed around collaboration between AI agents and human sleep experts (\Cref{fig:overview}). A conversational Interface Layer directs three specialist agents on behalf of the expert, who uses it to formulate research questions, launch runs on the cluster, review results, and re-enter structured, file-anchored feedback. The Hypothesis Agent, seeded by a broad direction from the expert, explores data-feasible hypotheses and returns a scored portfolio to curate; the Preprocessing Agent designs and validates feature-extraction pipelines on raw PSG before any full-cohort analysis; and the Execution Agent carries a selected hypothesis through staged statistical analysis to a report in which every reported number traces to an executable script. All components share a read-only data substrate spanning more than 50\,TB of raw PSG, in which validated signal-derived features are cached and reused, so that repeated analyses of tens of thousands of recordings do not require reprocessing the underlying signals for every hypothesis.

With the support of \method, we analysed roughly 124{,}000 PSG recordings across four cohorts through five case studies spanning three directions: how sleep physiology relates to disease risk, how it defines clinical phenotypes, and how sleep itself is organized and regulated. Two studies focused on \emph{disease risk}: whether coupling among the brain, heart, and peripheral systems is associated with incident neurodegenerative disease, and whether multidomain sleep-age residuals reflect current disease burden, future risk, or both. One addressed \emph{clinical phenotyping}: whether cortical arousal dynamics can reveal a clearer physiological signature of comorbid insomnia and sleep apnoea (COMISA), a contested category that conventional PSG summaries do not readily distinguish. Two examined \emph{mechanism}: the short-term regulation of rapid eye movement (REM) sleep and its relationship to the emergence of deep sleep, and whether the transient-oscillation organization of sleep microstructure is disrupted in narcolepsy type~1 (NT1). Each study began with a broad direction proposed by a sleep researcher and was refined through exploratory rounds with \method into a specific, data-feasible question before full-cohort analysis.

Across these five studies, the analyses reproduced established sleep physiology and identified patterns not previously evaluated at this scale: reduced sleep network-physiology coupling associated with incident Parkinson's and Alzheimer's disease diagnoses; a physiologically structured late-fusion sleep-age model whose disease associations generalized across cohorts; an obstructive sleep apnoea (OSA)-skewed intermediate COMISA phenotype characterized by prolonged post-arousal wakefulness and irregular arousal organization; a stronger association of subsequent REM-bout duration with inter-REM non-REM (NREM) duration than with intervening wake; and transient-oscillation disruptions in NT1.

\section*{\method overview}

\method is built around two principles. First, hypothesis generation is data-driven but shaped by expert priors: the expert sets the research direction, the Hypothesis Agent explores what the available data can actually support and returns scored candidate questions, and the expert selects which of them to advance. Second, expert authority is preserved at every irreversible step, so that hypothesis selection and full-cohort commitments occur only after explicit confirmation.

A typical project is a recurring loop between the expert and the system (\Cref{fig:overview}a). Through a chat-based \emph{Interface Layer} with persistent project memory, the expert brainstorms directions, launches runs, reviews reports and figures, and returns file-anchored feedback. From that interface the expert launches one of three specialist agents: the \emph{Hypothesis Agent}, which turns a broad direction into a scored portfolio of data-feasible candidate questions; the \emph{Preprocessing Agent}, invoked when a hypothesis requires a feature not yet extracted from raw signal; and the \emph{Execution Agent}, which carries a chosen hypothesis through staged analysis to a code-grounded report. Each specialist coordinates a small team of sub-agents that iterate until the work passes the agent's internal review. Expert feedback then re-enters the relevant agent as a first-class input, carrying the same authority as the original specification.

All four components operate against a shared, read-only data substrate that fixes both what data is exposed and how the agents reach it. Features are served through a schema-validated \emph{describe}, \emph{list}, and \emph{load} interface rather than prompt-embedded documentation that can drift from the underlying files. A single canonical-split function returns the authoritative partition of recordings into training, validation, and test sets, so that an agent cannot invent its own split. A shared \emph{Skills} library supplies the figure-style standards and sleep-science conventions common to every agent.

Three constraints keep the outputs grounded in the data. Candidate hypotheses must pass a data-feasibility check that rejects questions requiring variables that are neither preprocessed nor derivable from the raw signal. Feature-extraction pipelines are validated on a small sample of recordings and documented for expert review before any full-cohort extraction begins. Every reported number, figure, and table is bound to an executable script that re-derives it, with each citation resolved against its retrieved source. Across all three, an independent critic that cannot execute or alter the work it reviews audits each iteration. Supplementary~\Cref{fig:code_grounding_aging,fig:code_grounding_network} trace two reported quantities back through the code that produced them as representative examples. The Methods section describes the Interface Layer, the three specialist agents and their sub-agents, and these safeguards in full.


\section*{Data and research questions}

\method drew on four clinical and epidemiological PSG cohorts: the Stanford Sleep Clinic (SSC)~\citep{thapa2024sleepfm,kjaer2025stanford}, the Human Sleep Project (HSP)~\citep{li2026hsp}, the BioSerenity clinical network~\citep{hanif2023automatic, hanif2024associations, hanif2026use}, and the Central Disorders of Hypersomnolence (CDH) registry~\citep{stephansen2018neural, zhang2018national}, together comprising roughly 124{,}000 PSG recordings (Supplementary~\Cref{tab:cohort_demographics}). They differ in design, and these differences determine which questions each can address. SSC and HSP are large clinical PSG archives with linked electronic health records providing longitudinal follow-up and diagnoses coded with the International Classification of Diseases (ICD). They are the only cohorts in the substrate that support prospective analyses linking a single night of sleep to a diagnosis recorded years later, and because they were assembled independently at different institutions, SSC can serve as a discovery cohort and HSP as external validation. BioSerenity is a multi-site clinical network of comparable scale that lacks linked diagnostic codes but contains sleep questionnaires, respiratory measures, and symptom histories, making it suited to phenotyping questions whose group definitions rest on symptoms combined with physiology rather than on coded diagnoses. CDH is a smaller multi-site registry with harmonized diagnoses of the central disorders of hypersomnolence, including narcolepsy type~1, and therefore supports mechanistic questions in a population with a specific and well-characterized neurochemical lesion.

To support rapid hypothesis execution, we precomputed a curated feature library across the four cohorts, spanning sleep macrostructure and microstructure together with autonomic, respiratory, and brain-physiological coupling (Supplementary~\Cref{tab:features}).
The human-led features derive from validated community tools, including sleep staging from U-Sleep~\citep{perslev2021u}, cortical arousals from the Multimodal Arousal Detector~\citep{brink2020automatic}, apnoea events from the Apnoea-Based Event Detector~\citep{kjaer2026expert}, transient oscillations from DynamO~\citep{stokes2023transient}, and an established formulation of brain--heart coupling~\citep{Saibene2026BrainHeart}, alongside each cohort's curated clinical metadata.
The agent-led features were curated, adapted, or newly implemented by the Preprocessing Agent under expert review. Some are adaptations of existing community tools that the agent identified from the literature, pulled from source repositories, adapted for compatibility across cohorts, and exposed through the MCP interface. These include event-level microstructure such as spindles, slow waves, eye movements, and EEG artefact masks detected with Yet Another Spindle Algorithm (YASA)~\citep{vallat2021open}, and stage-stratified physiological coupling networks quantified by time-delay stability (TDS)~\citep{bashan2012network}. Others were designed by the agent directly from published methods, including R-peak detection, continuous oximetry, desaturation events, and heart-rate-variability (HRV) summaries~\citep{pukkila2026chronic}. Building on the original brain--heart coupling formulation, the agent further constructed a brain--EMG coupling measure on the same principle~\citep{Saibene2026BrainHeart}. All features were harmonized and post-hoc verified across cohorts.

The five areas of investigation were not fixed in advance of the data. Each began as a broad direction proposed by a sleep researcher from an open question in the field, was checked against what the substrate could actually support, and was then sharpened into a specific question through exploratory rounds between the investigator and \method before any full-cohort analysis was committed. The Hypothesis Agent triaged candidate questions for scientific merit and data feasibility, and the investigator used the Interface Layer to select which of them advanced (\Cref{fig:overview}b). The five case studies therefore reflect the intersection of expert research priorities with the particular strengths of the available cohorts.

Although the five studies address different disorders and timescales, they share a common setting: each asks what a single night of polysomnography reveals at a particular level of physiological organization. Those levels differ across the studies, spanning coupling between physiological systems, the domain composition of physiological ageing, the temporal organization of cortical arousals, the duration structure of cycling between NREM stages~1--3 (N1--N3) and REM sleep, and oscillatory events shorter than a scoring epoch. These questions group into three directions. Two case studies concern \emph{disease risk}, motivated by the observation that a single night of sleep carries signal predictive of disease years later without a shared framework for determining which features carry which risks. They ask whether coupling among brain, heart, and peripheral systems is associated with the risk of incident neurodegenerative disease, and whether multidomain sleep-age residuals are associated separately with prevalent disease burden and future disease risk. One concerns \emph{clinical phenotyping}, motivated by contested clinical classifications that conventional PSG summaries cannot resolve: whether the dynamics of cortical arousal can decompose COMISA into a clearer physiological signature. Two concern \emph{mechanism}, motivated by unresolved questions about how sleep is biologically organized: whether short-term REM-sleep regulation tracks preceding NREM sleep, how inter-REM NREM duration is distributed in relation to N3 emergence across the night, and whether the transient-oscillation organization of sleep microstructure is disrupted in NT1. The remainder of this paper presents the findings from each case study in turn.


\section*{Disorder-specific network-physiology coupling profiles during sleep}

\begin{mainfinding}
Neurological and neuropsychiatric disorders, particularly Parkinson's disease and Alzheimer's disease, showed distinct sleep-stage- and subnetwork-specific alterations in physiological coupling with moderate concordance across SSC and HSP.
In SSC, each 1\,SD decrease in N2 brain--brain coupling was associated with a higher risk of subsequent Parkinson's disease (HR 1.48, 95\% CI 1.31--1.67) and Alzheimer's disease (HR 1.38, 95\% CI 1.25--1.53) diagnoses.
\end{mainfinding}

\begin{figure}[!htbp]
    \centering
    \includegraphics[width=\linewidth]{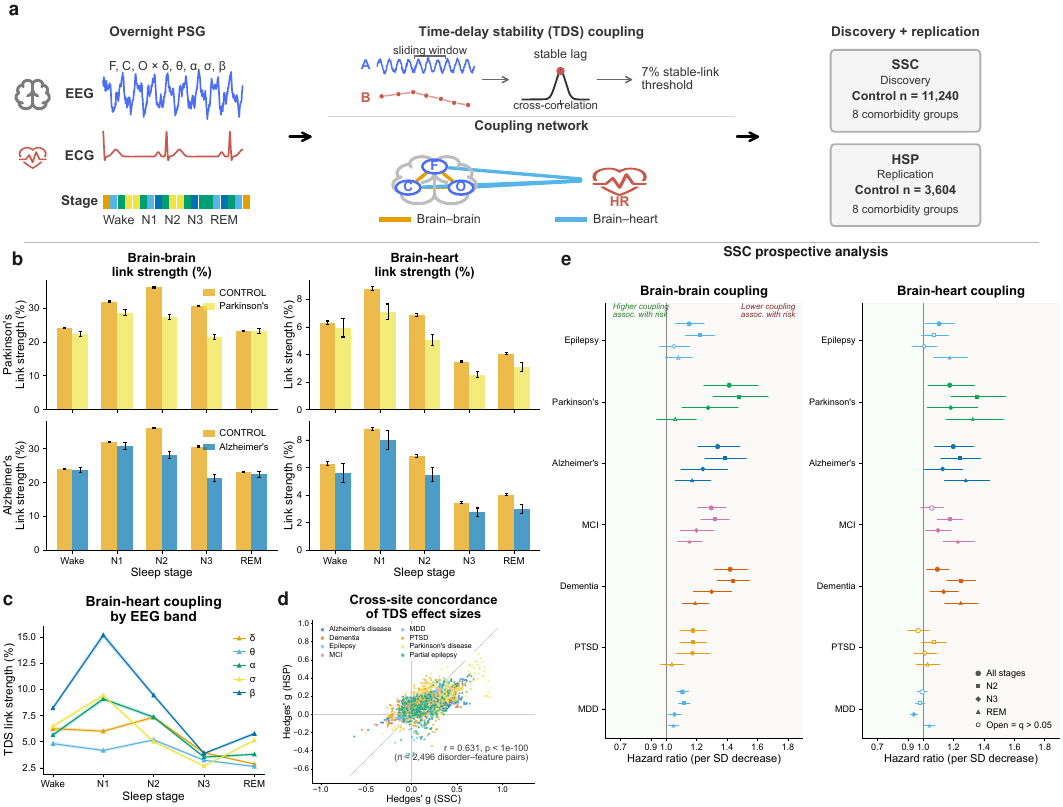}
    \caption{\textbf{Network-physiology coupling during sleep across neurological and neuropsychiatric disorders.}
    (a) Study design. EEG and ECG-derived heart-rate signals from overnight polysomnography were analysed within Wake, N1, N2, N3, and REM sleep using time-delay stability (TDS). TDS applies sliding-window cross-correlation to pairs of physiological signals and retains links for which the estimated time delay remains stable in more than 7\% of windows, producing subject-level brain--brain and brain--heart coupling networks for each sleep stage. Cross-sectional effect sizes were estimated in SSC (discovery; controls, $n=11{,}240$; eight disorder groups) and replicated in HSP (controls, $n=3{,}604$; the corresponding eight disorder groups).
    (b) Brain--brain and brain--heart TDS link strength across sleep stages in SSC controls and individuals with Parkinson's disease or Alzheimer's disease. Points and error bars show means and 95\% confidence intervals.
    (c) Brain--heart TDS coupling in SSC controls, stratified by sleep stage and EEG frequency band ($\delta$, $\theta$, $\alpha$, $\sigma$, and $\beta$).
    (d) Cross-site concordance of covariate-adjusted TDS effect sizes, shown as SSC versus HSP Hedges' $g$ across 2{,}496 disorder--feature pairs (Pearson $r=0.631$). Positive values indicate lower coupling in the disorder group than in controls, whereas negative values indicate higher coupling. Points in the upper-right and lower-left quadrants therefore show concordant effect directions across cohorts.
    (e) Prospective analysis in SSC. Points and error bars show hazard ratios and 95\% confidence intervals from Cox proportional-hazards models per 1\,SD decrease in brain--brain or brain--heart coupling, stratified by incident outcome and whole-night or sleep-stage-specific coupling summary. Prevalent cases were excluded separately for each outcome. Open markers denote associations with $q>0.05$ after false-discovery-rate (FDR) adjustment.}
    \label{fig:network_physiology}
\end{figure}

Sleep is a coordinated physiological state in which interactions between the brain and body continually reorganize across sleep stages. Neurological and neuropsychiatric disorders may therefore alter not only individual PSG signals, but also how different physiological systems interact over time. Consistent with this possibility, a recent foundation-model study of polysomnography found that cross-modal representations learned across electroencephalographic, respiratory, and other physiological signals outperformed single-modality representations on health-prediction tasks~\citep{kjaer2025stanford}. However, it remains unclear whether different disorders produce distinct alterations in physiological coupling during sleep, whether these patterns generalize across cohorts, and whether disrupted coupling precedes subsequent disease diagnosis.

To characterize these interactions, we used time-delay stability, a framework for identifying stable coupling between physiological signals~\citep{bashan2012network}. TDS calculates the cross-correlation between each pair of signals across successive sliding windows and identifies the time lag at which the correlation is maximal. Link strength was defined as the proportion of windows in which this lag remained stable, with a signal pair classified as significantly coupled when this proportion exceeded 7\%. These pairwise measurements were used to construct sleep-stage-specific physiological networks.

The analysis used SSC as the discovery cohort ($n=33{,}202$) and HSP as an independent replication cohort ($n=17{,}810$)~\citep{li2026hsp}; cohort selection and group counts are reported in Supplementary~\Cref{tab:np_cohort_flow}. In each cohort, the Control group was identified using PheCode-based criteria and was required to have no recorded history of neurological disorders (epilepsy, Parkinson's disease, dementia, Alzheimer's disease, or mild cognitive impairment [MCI]); neuropsychiatric disorders (major depressive disorder [MDD] or post-traumatic stress disorder [PTSD]); cardiovascular diseases (hypertension, atrial fibrillation, heart failure, acute myocardial infarction, or ischaemic heart disease); sleep disorders (insomnia or REM sleep behaviour disorder); or cerebrovascular events, including stroke or transient ischaemic attack. The Control groups comprised 11{,}240 SSC and 3{,}604 HSP participants. These participants were compared with eight neurological and neuropsychiatric disorder groups to determine whether different disorders exhibited shared or distinct coupling profiles and whether reduced coupling was associated with subsequent disease diagnosis.

The Control group provided a baseline for characterizing network physiology within each sleep stage. Within SSC, the brain--brain and brain--heart subnetworks showed distinct sleep-stage-dependent profiles. Brain--brain coupling increased from Wake to N1 (link strength: 24.1\% and 31.9\%, respectively), peaked during N2, declined during N3, and reached its lowest level during REM sleep. Brain--heart coupling followed a different profile, peaking during N1 before decreasing through N2 and reaching its minimum during N3 (link strength: 3.5\%) (\Cref{fig:network_physiology}b,c). Global brain--body coupling was also strongest during N2 and lowest during REM sleep. The attenuation of coupling from lighter NREM sleep to N3 was consistent with previous TDS findings~\citep{bashan2012network}. The feature landscapes of the SSC and HSP Control groups were highly correlated across 70 features ($r=0.979$, $p<10^{-48}$). Nevertheless, differences in absolute coupling remained between cohorts: brain--heart features showed small standardized differences, whereas brain--brain coupling was systematically higher in SSC than in HSP, particularly during N2 and N3.

To characterize disorder-specific network-physiology profiles, the Control group was compared separately with PheCode-defined groups for epilepsy, partial epilepsy, Parkinson's disease, Alzheimer's disease, dementia, MCI, PTSD, and MDD (Supplementary~\Cref{fig:np_supp_stage,fig:np_supp_volcano}). Analyses were conducted separately in SSC and HSP, with SSC serving as the discovery cohort and HSP as the replication cohort. Cross-sectional differences were quantified using Hedges' $g$ after adjustment for age, sex, and body mass index (BMI) in SSC and for age and sex in HSP, where positive values of $g$ indicated lower coupling in the disorder group than in the Control group.

In SSC, Parkinson's disease showed the broadest reduction in network coupling, with the largest effect observed for the percentage of significant frontal brain--brain links during N1 (Hedges' $g=0.86$). Alzheimer's disease, dementia, and MCI showed similar reductions in frontal--central brain--brain link strength during N1 (Hedges' $g=0.68$, $0.58$, and $0.51$, respectively). Epilepsy was associated with modestly lower frontal brain--brain coupling during N1 and N2, with the larger effect occurring during N1. Partial epilepsy showed its largest effect in the percentage of significant frontal brain--brain links during N2 (Hedges' $g=0.51$).

The neuropsychiatric groups showed more selective profiles. PTSD had the lowest proportion of FDR-significant features (23.4\%) and modest effect sizes (Hedges' $|g|=0.30$), with its largest effect observed for a reduction in the percentage of significant brain--EMG links during N2. In MDD, the largest overall effect was a reduction in brain--EOG coupling during N2. However, MDD differed from the other groups in showing predominantly higher brain--heart coupling relative to controls.

Cross-site replication was assessed by repeating the disorder comparisons in HSP and comparing the resulting effect-size patterns with those observed in SSC. Across 2{,}496 disorder--feature pairs, Hedges' $g$ values showed moderate concordance between cohorts (Pearson $r=0.631$;~\Cref{fig:network_physiology}d). Concordance was strongest for Parkinson's disease, dementia, and MDD, and weakest for PTSD. Because BMI was unavailable in HSP, the SSC analysis was repeated with and without BMI adjustment for five representative features across all eight disorder groups. The largest observed difference was for the percentage of significant frontal brain--brain links during N1 in Parkinson's disease (Hedges' $g_{\mathrm{with\,BMI}}=0.877$; Hedges' $g_{\mathrm{without\,BMI}}=0.908$).
Adjustment for BMI did not change the disorder rankings or which comparisons were statistically significant, indicating that BMI adjustment was unlikely to account for the cross-site agreement observed for the tested features.

To determine whether reduced coupling was associated with subsequent diagnoses, we fitted Cox proportional-hazards models to time-to-event data in SSC and HSP. For each outcome, participants with a diagnosis recorded before the PSG were excluded, and follow-up extended from the PSG date to the first recorded diagnosis or the last documented clinical contact. Models were adjusted for age, sex, and BMI in SSC and for age and sex in HSP. Hazard ratios estimated the change in risk associated with each 1\,SD decrease in coupling. In SSC, lower brain--brain coupling was associated with an increased risk of all seven incident outcomes examined, with the strongest associations generally observed during N2 sleep (\Cref{fig:network_physiology}e).

Lower brain--heart coupling was also associated with a higher risk of subsequent Parkinson's disease, Alzheimer's disease, and dementia diagnoses, primarily during N2, N3, and REM sleep. The association with Parkinson's disease was particularly pronounced (\Cref{fig:network_physiology}b,e). This finding is biologically plausible because progressive loss of cardiac noradrenergic sympathetic innervation is a characteristic feature of Parkinson's disease and can occur independently of nigrostriatal dopaminergic degeneration~\citep{candia2024framework}.

For N2 brain--brain coupling, each 1\,SD decrease was associated with a higher risk of subsequent Parkinson's disease (HR 1.48, 95\% CI 1.31--1.67) and Alzheimer's disease (HR 1.38, 95\% CI 1.25--1.53) diagnoses (\Cref{fig:network_physiology}e). The observed reduction in brain--brain coupling in these two neurodegenerative disorders is biologically plausible. One possible explanation is that neurotransmitter-specific deficits in dopamine in Parkinson's disease and acetylcholine in Alzheimer's disease, together with proteinopathy-related synaptic and interneuron impairment, disrupt the excitatory--inhibitory balance and structural connectivity required for coordinated oscillatory coupling across brain networks.

\begin{collaboration}
The investigator framed the initial question based on previous work on network-physiology coupling during sleep, and exploratory hypotheses were subsequently refined with \method in response to the primary findings. For each analytical step, the investigator specified the objective, required outputs, and scope constraints. Within these constraints, \method autonomously implemented the PheCode-based group definitions, extracted and summarized features from the precomputed TDSpy outputs, performed the effect-size analyses, assessed cross-site replication, fitted the Cox survival models, and generated the figures. Across four iterative rounds, feedback from the investigator and domain expert focused on clarifying the methodology, narrowing the analytical scope, and decomposing complex requests into sequential, single-objective tasks. \method incorporated this feedback into subsequent analyses. This process also revealed that broad, multi-part prompts could lead \method to prioritize the most prominent findings while overlooking other elements of the requested scope, motivating the use of narrower prompts with explicitly defined outputs.
\end{collaboration}


\section*{Multidomain polysomnographic sleep ageing reflects current disease burden and future disease risk}

\begin{mainfinding}
Physiologically structured late fusion outperformed unrestricted early fusion for PSG-based age prediction in the Stanford Sleep Clinic cohort and generalized more accurately to the Human Sleep Project cohort. Its bias-corrected age residual was associated with current disease burden and a reproducible pattern of incident cardiometabolic, renal, and respiratory disease. Disease-specific reweighting of the seven domain residuals did not improve held-out discrimination of the broad incident-disease composite.
\end{mainfinding}

\begin{figure}[!htbp]
    \centering
    \includegraphics[width=0.93\linewidth]{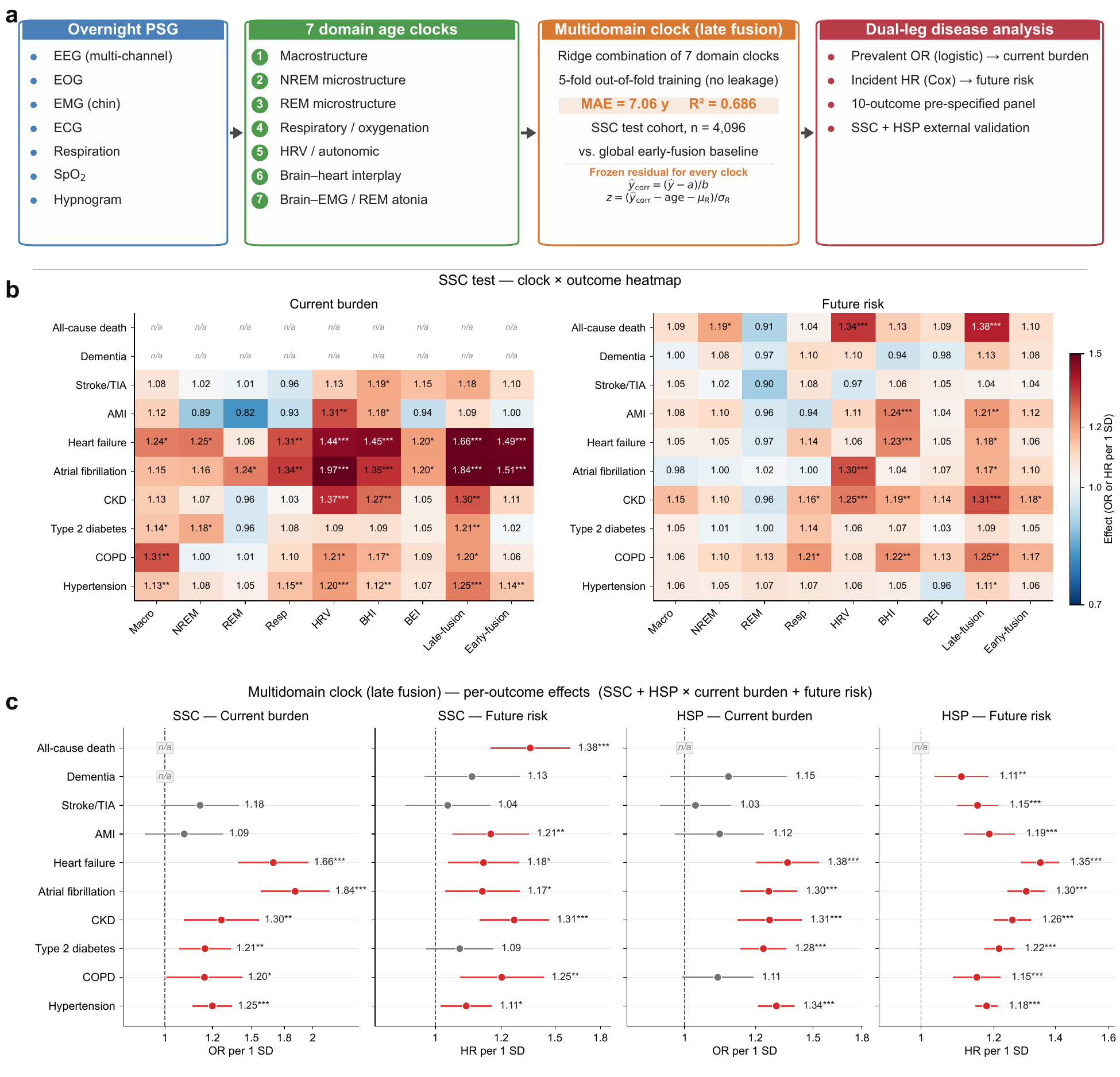}
    \caption{\textbf{Multidomain polysomnographic sleep-ageing residuals are associated with prevalent disease burden and incident disease risk.}
    (a) Analysis pipeline. PSG signals were represented by seven physiologically interpretable domain age clocks: macrostructure, NREM microstructure, REM microstructure, respiratory/oxygenation physiology, HRV/autonomic physiology, brain--heart interplay, and brain--EMG/REM-atonia. Predictions from the seven clocks were combined using a ridge late-fusion model trained on five-fold out-of-fold predictions. The resulting model achieved a mean absolute error (MAE) of 7.06 years and $R^2=0.686$ in the SSC test set ($n=4{,}096$ of the 4{,}120 test-split recordings, the remainder lacking required features; Supplementary~\Cref{tab:aging_cohort_flow}). Predicted ages were linearly bias-corrected, and residuals were standardized using parameters estimated from SSC training and validation recordings with an apnoea--hypopnoea index (AHI) $\leq 50$. These parameters were then fixed and applied to the SSC test set and HSP.
    (b) Clock-by-outcome heatmaps for the prespecified 10-outcome panel in SSC, shown for the seven domain clocks, the multidomain late-fusion model, and the global early-fusion baseline. Prevalent associations were estimated using logistic regression and represent disease burden at the time of PSG. Incident associations were estimated using Cox models after excluding prevalent cases for each outcome and represent future disease risk. Effects are reported per 1\,SD increase in the corresponding age residual.
    (c) Associations of the multidomain late-fusion residual with individual outcomes in SSC and HSP. Points and lines show effect estimates and 95\% confidence intervals; red denotes nominal significance and grey denotes $p\geq0.05$. Asterisks denote $^{*}p<0.05$, $^{**}p<0.01$, and $^{***}p<0.001$. \emph{n/a} indicates that prevalent mortality was not applicable, mortality follow-up was unavailable in HSP, or the prevalent-dementia model in SSC did not converge because only 27 cases were available. Incident associations were broadly consistent across outcomes available in both cohorts.}
    \label{fig:aging}
\end{figure}

Sleep-derived brain-age models can predict chronological age from PSG, and differences between predicted and chronological age have been associated with mortality, dementia, and other clinical outcomes~\citep{brink2022age, sun2019brain, sun2026machine}. One of the most widely used measures, the Brain Age Index, showed a mean absolute deviation of 7.6 years from chronological age among healthy participants in the Massachusetts General Hospital dataset and has also been associated with crystallized cognition~\citep{adra2023decoding}. Yet these associations do not establish what an elevated age residual represents: a general process of accelerated biological ageing, vulnerability within a particular physiological system, or changes resulting from established illness. Although shared mechanisms of ageing may increase susceptibility to disease, individual disorders also arise from organ-specific pathology, genetic predisposition, environmental exposures, and the consequences of existing illness~\citep{lopezotin2023hallmarks}. We therefore asked whether PSG-derived age residuals reflect a single global ageing axis or heterogeneous ageing trajectories across physiological systems, and whether their associations with disease capture existing disease burden, predict future disease risk, or both.

The analysis was developed in the SSC cohort and externally validated in a Harvard subsample of the Human Sleep Project cohort~\citep{li2026hsp}. After harmonized quality control, the analysis included 28{,}817 SSC recordings and 16{,}973 HSP recordings (Supplementary~\Cref{tab:aging_cohort_flow}). Seven domain-specific clocks captured sleep macrostructure, NREM and REM microstructure, respiratory and oxygenation physiology, HRV and autonomic physiology, brain--heart interplay, and brain--EMG and REM-atonia physiology. A ridge late-fusion model combined five-fold out-of-fold predictions from these clocks and achieved the best age-prediction performance on the held-out SSC test set, with an MAE of 7.06 years and $R^2=0.686$. By comparison, the strongest single-domain clock achieved an MAE of 8.05 years, and the unrestricted global early-fusion model achieved an MAE of 7.33 years (\Cref{fig:aging}a; Supplementary~\Cref{fig:aging_supp}a). Late fusion reduced MAE relative to early fusion by 0.27 years in paired-bootstrap analysis (95\% CI, 0.13--0.42 years), supporting the value of combining predictions across physiologically defined domains rather than directly aggregating all features. For each clock, predicted ages were corrected for age-related bias, and the resulting age residuals were standardized using parameters estimated from SSC training and validation recordings with AHI $\leq 50$. All model and correction parameters were then frozen before evaluation in the SSC test set and external validation in HSP.

The principal disease analysis evaluated a prespecified panel of 10 outcomes in parallel prevalent and incident analyses, where applicable. The panel included all-cause mortality, dementia, cardiovascular diseases, chronic kidney disease (CKD), type~2 diabetes, and chronic obstructive pulmonary disease (COPD). Logistic regression was used to test whether the age residual was associated with disease already present at the time of PSG. For each incident analysis, participants with the corresponding disease at baseline were excluded, and Cox regression was used to assess subsequent disease risk. This parallel design allowed each outcome to be interpreted as reflecting current disease burden, future disease risk, both, or neither.

The associations did not support a uniform, global signal in which sleep-age residuals predicted all outcomes. Instead, they followed a coherent pattern across cardiometabolic, renal, and respiratory disease (\Cref{fig:aging}b). The late-fusion and HRV residuals showed the broadest associations, whereas the brain--heart interplay residual was particularly associated with cardiovascular outcomes. In incident analyses, the late-fusion residual was associated with all-cause mortality, CKD, COPD, acute myocardial infarction, heart failure, atrial fibrillation, and hypertension; associations with dementia and stroke or transient ischaemic attack were weaker in SSC. Prevalent associations were strongest for heart failure, atrial fibrillation, CKD, type~2 diabetes, COPD, and hypertension. Together, these patterns suggest that the residual reflects a combination of shared physiological vulnerability, disease-specific dysfunction, and changes associated with established disease, rather than a single causal pathway of ageing.

External validation provided further support for the physiologically structured late-fusion approach. When applied to HSP without retraining, age-prediction performance declined, potentially reflecting cohort shift; nevertheless, late fusion generalized better (MAE 8.83 years, $R^2=0.494$) than the global early-fusion baseline (MAE 9.52 years, $R^2=0.315$). Associations between the late-fusion residual and individual outcomes were also broadly consistent across SSC and HSP (\Cref{fig:aging}c). Incident associations were particularly consistent, whereas prevalent associations varied more between cohorts, possibly owing to differences in referral patterns, baseline disease burden, clinical coding, and covariate availability. Disease-tuned models were evaluated as secondary risk scores rather than as alternative ageing clocks. For the broad disease composite, disease-specific reweighting of the domain residuals provided no improvement in discrimination over the age-derived late-fusion residual in either SSC ($\Delta C=-0.005$ vs.\ $+0.002$) or HSP ($\Delta C=+0.004$ vs.\ $+0.005$), relative to a demographic baseline model.

A phenome-wide association study (PheWAS) extended these findings beyond the prespecified 10-outcome panel (Supplementary~\Cref{fig:aging_supp}b). In SSC, the late-fusion residual was associated at FDR significance with 39 incident and 80 prevalent PheCodes. Associations were concentrated in disease categories related to the primary panel, including circulatory, endocrine/metabolic, renal, genitourinary, and respiratory disorders. Among the SSC-significant PheCodes that could be evaluated in HSP, effect directions were concordant for 100\% of incident and 94\% of prevalent associations. This breadth suggests that the multidomain PSG age residual captures clinically meaningful physiological dysfunction, whereas its concentration within particular organ systems argues against interpreting it as a universal mediator of disease.

To assess whether the findings depended on a particular agent trajectory, the complete study was independently repeated three times from the same high-level prompt (Supplementary~\Cref{tab:aging_replication}). Although the agent developed different feature spaces across runs, leading to variation in age-prediction accuracy, the main findings were consistently reproduced: associations spanning cardiometabolic, renal, respiratory, and autonomic disease; generalization of incident PheWAS associations; and limited incremental benefit from disease-specific reweighting of the age residuals.

The value of this case study lay not only in the model development but in how it refined the scientific question. Rather than treating a PSG-derived age gap as a universal measure of biological ageing~\citep{brink2022age, sun2019brain, sun2026machine, adra2023decoding}, the analysis decomposed sleep ageing into physiologically defined domains and showed that late fusion yielded a more interpretable representation that generalized better across cohorts. The resulting residual does not have a single biological meaning: depending on the outcome, it may reflect shared age-related vulnerability, disease-specific pathophysiology, consequences of established disease, or a combination of these processes~\citep{lopezotin2023hallmarks}. Relative to previous sleep-age studies, the principal advance lies in combining physiologically structured multimodal age modelling with parallel evaluation of prevalent disease burden and incident disease risk, followed by external validation. Together, these analyses revealed a reproducible pattern of associations centred on cardiometabolic, renal, respiratory, and autonomic dysfunction, rather than a universal signal across all diseases.

\begin{collaboration}
The investigator established the initial direction based on prior literature, and the study was subsequently expanded to compare multiple global modelling strategies as the agent's exploratory runs surfaced additional possibilities. Before scaling the analysis, the investigator specified the seven physiologically interpretable feature domains, the parallel prevalent and incident disease analyses, the prespecified 10-outcome clinical panel, and the SSC-to-HSP external-validation design. Within this framework, \method engineered the features for each domain; trained the seven domain clocks, the multidomain late-fusion clock, and the global early-fusion baseline; conducted both disease analyses; and generated all figures. Methodological contributions flowed in both directions: \method proposed the age bias-correction procedure that was subsequently adopted into the pipeline and independently initiated sensitivity analyses designed to challenge the main findings. Across three rounds of structured feedback through the Interface Layer, the investigator requested additional features, clearer reporting of data exclusions, and figure-level revisions. The agent incorporated each request and, where the results warranted further investigation, added further sensitivity analyses.
\end{collaboration}

\section*{Arousal dynamics reveal an obstructive sleep apnoea-skewed intermediate phenotype in COMISA}

\begin{mainfinding}
Within the sleep-microstructure feature space examined here, comorbid insomnia and sleep apnoea (COMISA) more closely resembled obstructive sleep apnoea (OSA) than insomnia. Relative to OSA, COMISA was characterized by longer post-arousal wakefulness and greater within-night irregularity in arousal organization rather than greater respiratory disturbance. This suggests that, although respiratory events may trigger cortical arousals similarly in OSA and COMISA, individuals with COMISA have greater difficulty re-establishing stable sleep following arousal. Insomnia showed minimal separation from the Control group. Together, these findings identify cortical arousal dynamics as a complementary dimension of COMISA physiology beyond conventional respiratory measures.
\end{mainfinding}

\begin{figure}[!htbp]
    \centering
    \includegraphics[width=0.95\linewidth]{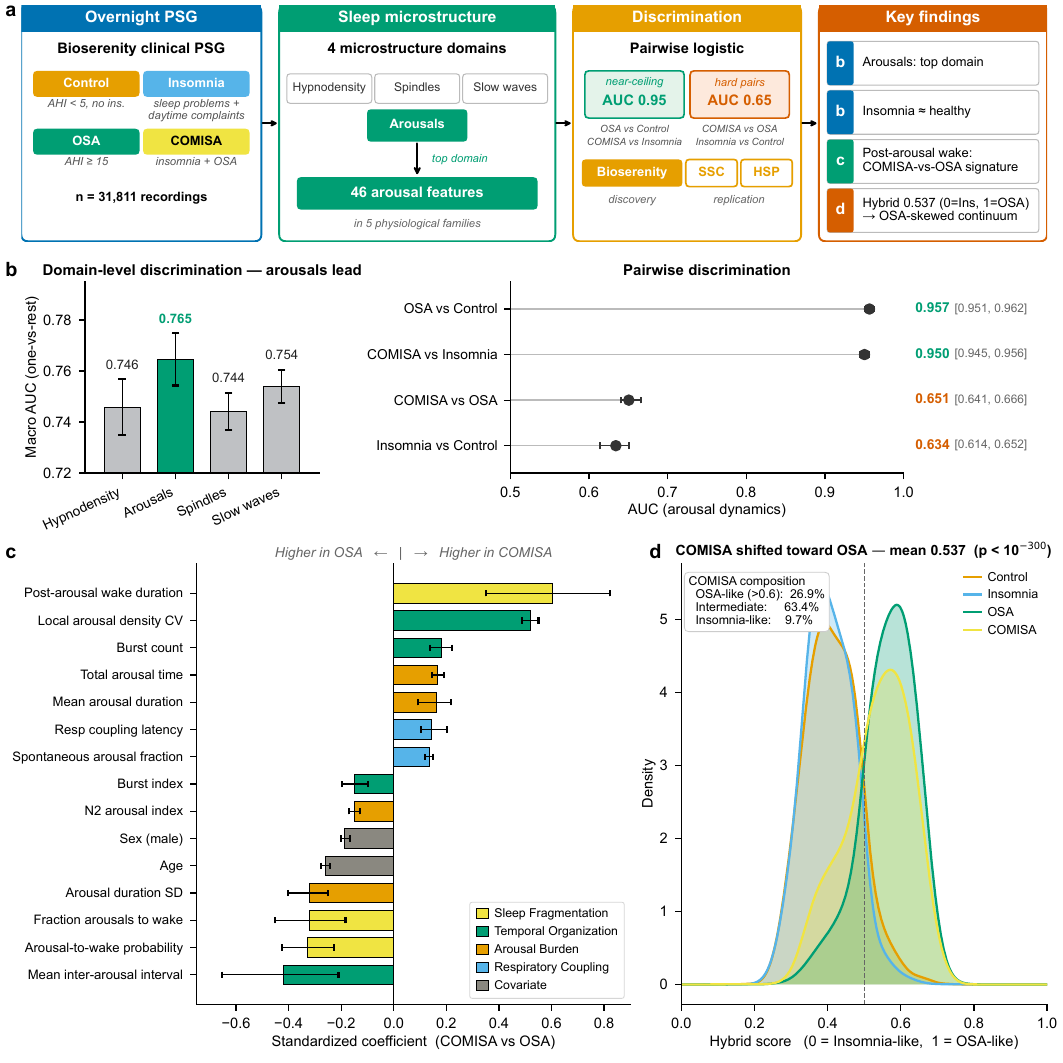}
  \caption{\textbf{Arousal dynamics characterize COMISA as an OSA-skewed intermediate phenotype.}
(a) Case-study schematic. Clinical polysomnography from the BioSerenity cohort was analysed across four diagnostic groups (Control, insomnia, OSA, COMISA; inset definitions). Four sleep-microstructure domains were first screened on $n_1 = 31{,}811$ recordings, followed by arousal dynamics with 46 features across four physiological families on $n_2 = 35{,}964$ recordings. Pairwise logistic classification revealed differences in diagnostic separability, with some comparisons classified with high accuracy (area under the receiver operating characteristic curve [AUC] $\approx 0.95$) and others proving substantially more challenging (AUC $\approx 0.65$). Three of four principal claims replicated across the SSC and HSP cohorts.
(b) Discrimination summary. \emph{Left}: macro-averaged one-vs-rest AUC per single sleep-microstructure domain (hypnodensity, arousals, spindles, slow waves); cortical arousals are the most informative single domain (0.765 versus 0.744--0.754 for the others). \emph{Right}: pairwise arousal-dynamics AUC with 95\% CIs: OSA versus Control (0.957 [0.951, 0.962]) and COMISA versus insomnia (0.950 [0.945, 0.956]) are near-ceiling, whereas COMISA versus OSA (0.651 [0.641, 0.666]) and insomnia versus Control (0.634 [0.614, 0.652]) are considerably harder.
(c) Covariate-adjusted standardized coefficients distinguishing COMISA from OSA, coloured by arousal-feature domain; positive values denote features elevated in COMISA. Longer post-arousal wakefulness and more irregular arousal organization (local arousal-density coefficient of variation) are the strongest COMISA-leaning signals.
(d) Distribution of a hybrid score (0 = insomnia-like, 1 = OSA-like) for each group; the COMISA distribution is shifted towards the OSA end (mean 0.537; $p < 10^{-300}$ versus 0.5), comprising 26.9\% OSA-like, 63.4\% intermediate, and 9.7\% insomnia-like recordings.}
  \label{fig:comisa}
\end{figure}

Obstructive sleep apnoea and insomnia are among the most common sleep disorders and frequently co-occur~\citep{icsd}. This combined phenotype, termed comorbid insomnia and sleep apnoea, affects an estimated 30--50\% of individuals with either disorder and is associated with greater daytime impairment, elevated cardiometabolic risk, and poorer treatment outcomes than either condition alone~\citep{Sweetman2017, Sweetman2023, Lechat2022}. Despite its clinical importance, the nocturnal physiology underlying COMISA remains poorly understood, with prior studies limited to small samples~\citep{sweetman2021bi, brooker2023obstructive}. A dominant model conceptualizes COMISA as a hybrid disorder arising from the interaction between insomnia-related hyperarousal and OSA-driven respiratory instability~\citep{sweetman2021bi}. However, it remains unclear whether COMISA represents a balanced integration of these mechanisms or is physiologically dominated by one constituent process.

There remains debate over whether COMISA is characterized by a reduced arousal threshold, meaning a greater propensity to awaken in response to an apnoeic event~\citep{brooker2023obstructive, yanagimori2023respiratory}. Conventional PSG metrics, including sleep efficiency, sleep-stage composition, and the AHI, provide only coarse summaries of sleep and respiration and are ill suited to addressing this question~\citep{hanif2026use}. A recent meta-analysis of 16 studies found that patients with COMISA had a higher microarousal index, higher AHI, lower minimum SpO$_2$, longer wake after sleep onset, and lower sleep efficiency than patients with OSA alone~\citep{cotrik2026physiopathological}. Patients with COMISA also showed longer awakenings ($\geq 5$ minutes), with significantly more wake time attributable to these extended awakenings than patients with OSA alone (43.3 versus 25.5 minutes, $p<0.001$)~\citep{cotrik2026physiopathological}. Multi-night home monitoring further showed that patients with COMISA had greater night-to-night variability in sleep-onset latency, sleep efficiency, and awakening duration, a pattern that cannot be captured by single-night PSG~\citep{cotrik2026physiopathological}. Together, these limitations make conventional PSG metrics poorly suited to capturing the temporal dynamics of cortical arousals, which mediate interactions among respiratory events, autonomic activation, and sleep fragmentation~\citep{de2018dynamic}.

To investigate this question, two analytic samples were derived from the BioSerenity PSG cohort (Supplementary~\Cref{tab:comisa_cohort}). We first screened four sleep-microstructure domains, namely cortical arousals, sleep spindles, slow waves, and hypnodensity-derived sleep-stage uncertainty, using 31{,}811 PSG recordings from the Control, insomnia, OSA, and COMISA groups (\Cref{fig:comisa}a). Two findings emerged. First, insomnia showed limited objective separation from the Control group, consistent with previous reports of limited reproducible PSG abnormalities in chronic insomnia~\citep{baglioni2014sleep, stephan2023reconsidering}. Second, cortical arousals were the most informative feature domain, with measures of arousal frequency, timing, and organization outperforming spindle-, slow-wave-, and staging-derived metrics in distinguishing among the diagnostic groups. Cortical arousals achieved the highest single-domain discrimination, with a macro-averaged AUC of 0.765, compared with 0.744--0.754 for the other domains (\Cref{fig:comisa}b; Supplementary~\Cref{fig:comisa_supp}a,c).

We then analysed a larger set of 35{,}964 PSG recordings for which reliable arousal detection was available, extracting 46 features describing arousal burden, respiratory coupling, temporal clustering, and post-arousal sleep--wake trajectories. Respiratory-linked arousal dynamics strongly distinguished groups that differed in sleep-disordered breathing. Pairwise logistic-regression models distinguished OSA from the Control group with an AUC of 0.957 (95\% CI 0.951--0.962) and COMISA from insomnia with an AUC of 0.950 (95\% CI 0.945--0.956) (\Cref{fig:comisa}b). In contrast, COMISA was substantially more difficult to distinguish from OSA (AUC 0.651, 95\% CI 0.641--0.666), as was insomnia from the Control group (AUC 0.634, 95\% CI 0.614--0.652). Nevertheless, the differences between COMISA and OSA were concentrated in arousal features beyond conventional measures of respiratory-event burden. The strongest discriminative features captured post-arousal wakefulness and the temporal organization of arousal events across the night (\Cref{fig:comisa}c). Relative to OSA alone, COMISA was characterized by longer wakefulness after arousal and greater irregularity in arousal timing. These findings are consistent with a recent study of 72 patients with COMISA and 72 patients with OSA, which found greater autonomic responses and longer arousals in COMISA~\citep{wulterkens2024heart}.

Arousal dynamics also provided information beyond conventional sleep and respiratory measures: adding these features to standard PSG indices improved diagnostic discrimination across all comparisons ($\Delta$AUC range: +0.042--0.123). To characterize the position of COMISA along the insomnia--OSA arousal spectrum, we computed a Mahalanobis-based hybrid score using the 15 features with the largest group effects. Features were z-scored, and the insomnia and OSA group centroids were estimated using a regularized pooled covariance matrix. Each recording's score was defined by its relative Mahalanobis distance to the two centroids, ranging from 0 (insomnia-like) to 1 (OSA-like). COMISA recordings formed a largely continuous distribution rather than separating into discrete insomnia- and OSA-dominant clusters (\Cref{fig:comisa}d; Supplementary~\Cref{fig:comisa_supp}b). The distribution was shifted towards the OSA end of the spectrum (mean hybrid score: 0.537; $p<0.001$ relative to the midpoint of 0.5), with 26.9\% of recordings classified as OSA-like, 63.4\% as intermediate, and 9.7\% as insomnia-like. These findings do not contradict previous reports that insomnia and OSA can interact synergistically to affect conventional respiratory measures and clinical outcomes~\citep{dai2026interactions}. Rather, they indicate that, within the sleep-microstructure feature space examined here, COMISA was more closely aligned with OSA than with insomnia, while its differences from OSA were concentrated in post-arousal wakefulness and the temporal organization of arousals.

To assess generalizability, the domain-screening and arousal-focused analyses were repeated in the independent SSC and HSP cohorts using the same diagnostic definitions and feature-extraction pipelines (\Cref{fig:comisa_supp}). Three findings were consistent across cohorts: insomnia showed minimal separation from the Control group, cortical arousals remained the most informative microstructure domain, and the residual distinction between COMISA and OSA was concentrated in post-arousal wakefulness rather than respiratory burden. This pattern is also consistent with a smaller study reporting increased cardiorespiratory responses to apnoeas during ``subcortical arousals'' in COMISA compared with OSA~\citep{wulterkens2024heart}. The precise positioning of COMISA along the insomnia--OSA spectrum was more cohort-dependent. BioSerenity placed COMISA at an OSA-skewed intermediate position, SSC found it largely indistinguishable from OSA, and HSP showed greater phenotypic polarization and evidence of sleep disruption exceeding that observed in either constituent disorder. The HSP finding is consistent with a previous large-scale study showing that insomnia and OSA can interact synergistically to amplify clinical symptoms and selected conventional PSG abnormalities, including respiratory-disturbance and arousal-related measures~\citep{dai2026interactions}. Respiratory-axis classification was also weaker in the replication cohorts than in BioSerenity. Despite these quantitative differences, all three cohorts supported the same broad interpretation: within the sleep-microstructure feature space examined here, COMISA was generally more closely aligned with OSA than with insomnia, although synergistic interactions between the two disorders may emerge in other physiological domains.

Together, these findings show that conventional PSG metrics do not fully capture the physiological heterogeneity of COMISA. Measures derived from cortical arousals provide complementary information and may improve physiological stratification beyond standard respiratory indices.

\begin{collaboration}
The investigator framed the initial direction based on prior work on the insomnia--apnoea overlap, while \method surveyed the available physiological features and cohorts and proposed candidate domains through which the question could be addressed. The investigator then fixed the scientific question, the four diagnostic labels (Control, insomnia, OSA, and COMISA), and the training, validation, and test partitions. Within these constraints, \method designed and executed the computational analyses, selected the modelling approaches, evaluated the features, performed cross-cohort validation, and generated all figures. Across four iterative rounds, investigator feedback refined the sub-hypotheses, clarified the analytical scope, and focused subsequent analyses on arousal dynamics as the most informative domain. This process also showed that shorter prompts with a single objective consistently produced more focused and reproducible outputs than compound, multi-part prompts.
\end{collaboration}

\section*{REM-bout duration is more strongly associated with NREM than wake, and a valley in inter-REM NREM duration precedes N3 emergence}

\begin{mainfinding}
Within inter-REM intervals, REM-bout duration was more strongly associated with NREM duration than with intervening wake, although both associations were significant. This relationship, previously reported in rodents and small human samples, was reproduced across large clinical cohorts, including individuals with narcolepsy type~1. Inter-REM NREM duration was also bimodally distributed, with the 20--25\,min valley closely preceding the first N3 epoch, suggesting a putative inhibition window.
\end{mainfinding}

\begin{figure}[!htbp]
    \centering
    \includegraphics[width=0.91\linewidth]{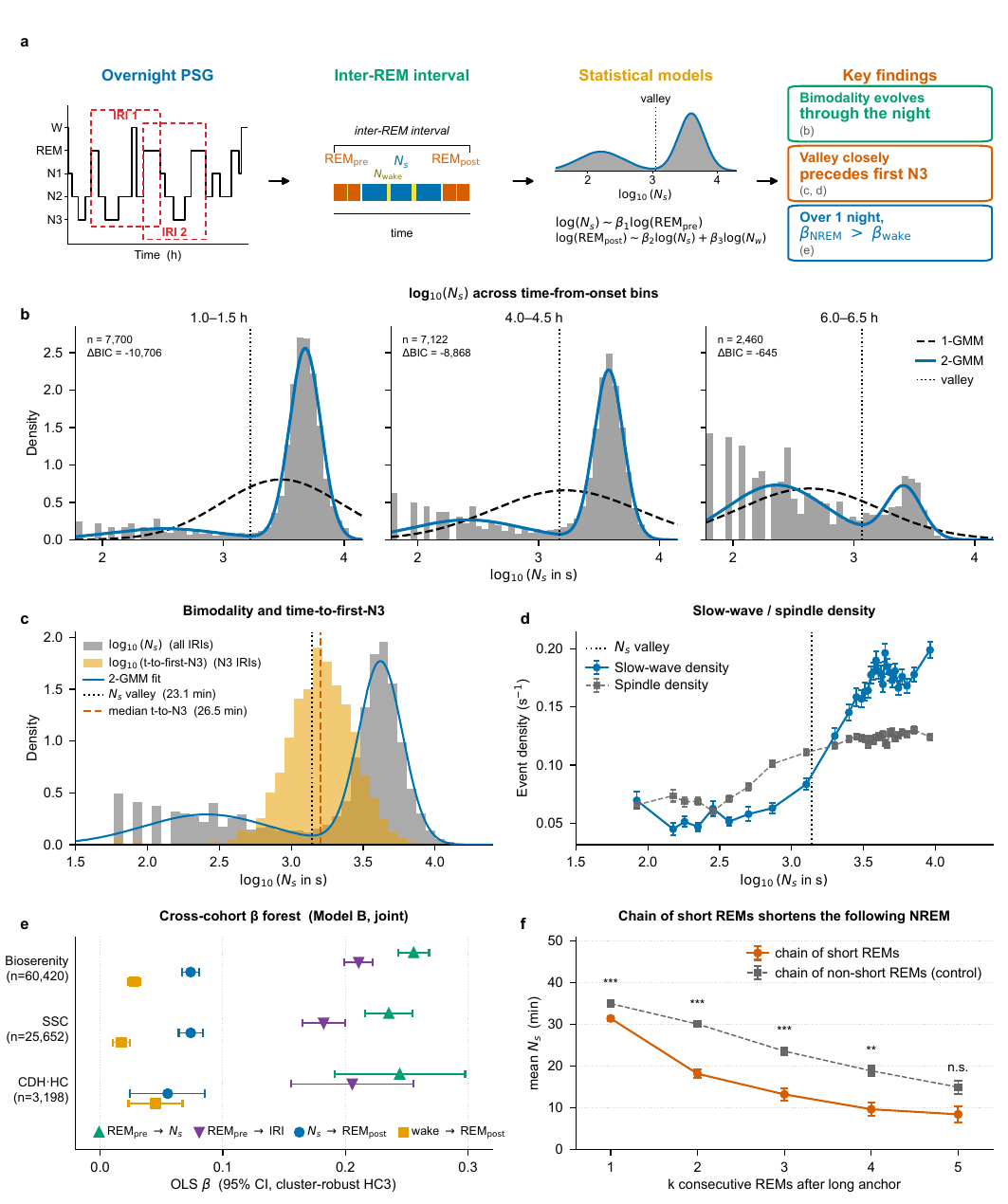}
    \caption{\textbf{Inter-REM NREM-duration bimodality, N3 emergence, and REM-bout associations across cohorts.}
    (a) Analysis schematic. 30-second hypnograms were generated by majority vote over 1-second hypnodensity epochs, and inter-REM intervals (IRIs) were extracted, including preceding REM bout ($\mathrm{REM_{pre}}$), inter-REM interval comprising NREM time ($N_s$) and intervening wake ($N_{\mathrm{wake}}$), and subsequent REM bout ($\mathrm{REM_{post}}$).
    (b) Distributions of $\log(N_s)$ across 30-min time-from-sleep-onset bins, showing changes in bimodality across the night.
    (c) Two-component Gaussian fit of $\log(N_s)$ and distribution of time to first N3. The $N_s$ valley (23.1\,min) precedes the median time to first N3 (26.5\,min).
    (d) Slow-wave and spindle densities across $\log(N_s)$; the $N_s$ valley coincides with the increasing slow-wave-density regime.
    (e) Covariate-adjusted ordinary least squares (OLS) coefficients relating $\mathrm{REM_{pre}}$ to subsequent $N_s$ and IRI total duration, and $N_s$ and $N_{\mathrm{wake}}$ to $\mathrm{REM_{post}}$ across three cohorts. $N_s$ was more strongly associated with $\mathrm{REM_{post}}$ than intervening wake was.
    (f) Consecutive short REM bouts lead to progressively shorter subsequent NREM durations. The orange curve shows the final IRI NREM duration after a long REM bout followed by $k = 1, 2, 3, 4$ short REM bouts (before a new long REM bout appears); the blue curve shows the corresponding control case, in which the long REM bout is followed by $k = 1, 2, 3, 4$ non-short REM bouts.}
    \label{fig:rem_regulation}
\end{figure}

While the homeostatic regulation of NREM-sleep pressure has been modelled with quantitative success~\citep{borb1999sleep,rajdev2013unified} and partially mapped onto identifiable molecular substrates~\citep{mongrain2025revisiting}, the homeostatic regulation of REM-sleep pressure (HREMSP) remains incompletely characterized~\citep{park2020neural}. Benington and Heller proposed that, over the course of a single night, HREMSP accumulates primarily during NREM sleep and dissipates during REM sleep~\citep{benington1994rem,benington1994does}. However, subsequent REM-deprivation studies support the idea that HREMSP also accumulates during wakefulness, motivating a distinction between short-term regulation of the NREM--REM cycle within a sleep period and long-term regulation of daily REM-sleep amount~\citep{ocampo2000homeostasis,franken2002long}. Furthermore, REM-sleep deprivation induces REM rebound in suprachiasmatic-nucleus-lesioned animals lacking overt circadian rhythmicity in the absence of zeitgebers, supporting a homeostatic component independent of circadian gating~\citep{wurts2000circadian}.

Recent studies have reported that NREM duration within inter-REM intervals (IRIs) follows a bimodal distribution in both rodents~\citep{ginsberg2024predictive} and humans~\citep{akhavan2026data}, consistent with earlier sleep-interruption studies suggesting a temporal relationship between NREM duration and REM-sleep occurrence~\citep{takeuchi2002elicitation,sasaki1993effects}. Here we tested whether this bimodality is a robust feature of human sleep architecture and examined its temporal relationship to the emergence of N3 sleep. We then assessed whether subsequent REM-bout duration was more strongly associated with NREM duration within the IRI than with intervening wakefulness, and whether prior REM-bout duration influenced the subsequent IRI and the NREM sleep within it. These relationships were examined across large clinical cohorts, including individuals with narcolepsy type~1, a disorder caused by orexin deficiency in which REM propensity is greatly enhanced~\citep{mignot2021sleep}.

The primary analysis comprised 24{,}409 BioSerenity night recordings comprising 74{,}079 IRIs, of which 60{,}420 had complete demographic and AHI covariate data. Because segmentation choices affect bout and sleep-cycle definitions~\citep{blume2021sleepcycles,ginsberg2024predictive}, sensitivity analyses covered 81 REM/NREM segmentation configurations, were stratified by wake-after-sleep-onset (WASO) burden, and re-evaluated the primary findings for each successive sleep cycle.
The analysis was extended to SSC (21{,}018 recordings, 69{,}502 IRIs, 25{,}652 with all demographic and AHI covariates) and the control subgroup of the CDH registry (1{,}181 recordings, 3{,}198 IRIs with all demographic and AHI covariates); the NT1 arm of CDH (501 recordings eligible for analysis, 2{,}433 IRIs with all demographic and AHI covariates) tested whether the same architecture held under disordered REM regulation.
All covariate-adjusted regression models included age, sex, BMI, and AHI.

The distribution of $\log(N_s)$ (log duration of NREM in the IRI) within IRIs was bimodal: a two-Gaussian mixture fitted the data substantially better than a unimodal Gaussian (\Cref{fig:rem_regulation}b; difference in Bayesian information criterion, $\Delta\mathrm{BIC} = -73{,}286$), a pattern that held across cohorts and across the 81 segmentation configurations.
Its shape evolved across the night: IRIs with short NREM duration were improbable shortly after sleep onset, became progressively more probable with time asleep, and bimodality became less distinct towards morning (\Cref{fig:rem_regulation}b). The valley threshold of the distribution (23.1\,min in BioSerenity) closely preceded the median duration at which the first N3 epoch appeared within the NREM bout (26.5\,min), and coincided with the rising regime of slow-wave density. This was observed in all three cohorts (\Cref{fig:rem_regulation}c,d, Supplementary~\Cref{fig:rem_sleep_supp}a,c).

Across the three cohorts, we tested the previously reported association between the duration of a REM bout and the duration of the subsequent IRI NREM~\citep{benington1994does}. The log duration of the REM bout preceding an IRI ($\mathrm{REM_{pre}}$) was associated with log total IRI duration. The association was stronger with $\log(N_s)$ (\Cref{fig:rem_regulation}e; covariate-adjusted OLS on the BioSerenity dataset with cluster-robust standard errors clustered at the recording level; CI derived from the cluster-robust covariance; $\mathrm{REM_{pre}}$ OLS coefficient for $\log_{10}(N_s)$ was 0.2477, SE 0.0064, $q < 0.001$). $\log(N_s)$ within an IRI was more strongly associated with $\log(\mathrm{REM_{post}})$ than was log wake duration in the same interval, though both were significant (\Cref{fig:rem_regulation}e; covariate-adjusted, $\beta_{\log_{10}(N_s)} = 0.0738$, SE 0.0035, Shapley partial $R^2$: 0.028, versus $\beta_{\log_{10}(N_\text{wake}+1)} = 0.0276$, SE 0.0026; Shapley partial $R^2$: 0.011, both $q < 0.001$). Finally, following successive short REM bouts, subsequent NREM durations were progressively shorter than after a long REM bout followed by non-short bouts (\Cref{fig:rem_regulation}f; thresholds: long REM bout $\geq 10$\,min, short REM bout $\leq 5$\,min; Mann--Whitney $U$ test, significant for $k = 1$ to $4$). Sensitivity analyses showed that this finding was robust to the choice of short-bout threshold.

In the NT1 subgroup, the associations shown in~\Cref{fig:rem_regulation}e and the across-night bimodality of $\log(N_s)$ were preserved, although the global valley threshold appeared earlier, at 18.7\,min. Splitting patients by sleep-onset REM period (SOREMp$+$, first transition to REM below 15\,min, aligned with American Academy of Sleep Medicine [AASM] recommendations, versus SOREMp$-$) did not reveal any difference in bimodality dynamics (Supplementary~\Cref{fig:rem_sleep_supp}b).

Three findings emerged consistently across cohorts and sensitivity configurations. First, IRI NREM duration was robustly bimodal, with a valley preserved across cohorts (23.1\,min for BioSerenity, 23.3\,min for the CDH control subgroup, and 21.4\,min for SSC).
Across the night, the IRI NREM duration distribution drifted towards more permissive short-bout regimes as sleep progressed.

Second, this valley coincided with the rising regime of slow-wave density and closely preceded the first N3 epoch. This temporal association is compatible with, but does not establish, a transient inhibition of the NREM-to-REM transition that would favour slow-wave consolidation. Because the pattern persisted into the late night when N3 is rare, it appears to be independent of the actual occurrence of N3. The observed pattern was also inconsistent with the proposed self-inhibition of REM sleep at its termination~\citep{le2021asymmetrical}, as early NREM-to-REM transitions are more probable than transitions within the 20--25\,min window throughout the night. In NT1 the same associations and bimodality held, with the valley and first N3 epoch appearing earlier than in controls.

Third, across several large clinical cohorts and in a population with disordered REM regulation, the reported stronger association of REM duration with NREM duration compared with wake during the IRI held~\citep{benington1994does}. Over a single night, the association of $\mathrm{REM_{post}}$ with IRI NREM duration was stronger than with IRI wake duration. Yet both were significant, which is compatible with the short- and long-term processes of HREMSP as suggested by Franken~\citep{franken2002long}. Moreover, because wake time (16\,h) typically exceeds NREM time (6.5\,h) over 24\,h by almost threefold, the net contributions of wake and NREM could balance out to end up becoming equivalent over a full day. While the $\mathrm{REM_{pre}}$--$N_s$ association is well established across species, this association between IRI NREM duration and subsequent REM-bout duration has not been reported in earlier rodent and human studies~\citep{benington1994rem, park2021probabilistic, barbato1998homeostatic}. The effect we resolve is small, suggesting stronger coupling between $\mathrm{REM_{pre}}$ and $N_s$ than between $N_s$ and $\mathrm{REM_{post}}$, and may have been beyond the reach of much smaller cohorts. A direct test would require an independent marker, such as the rate of REM-onset attempts under selective deprivation~\citep{franken2002long}.

The extent to which these associations inform HREMSP itself warrants caution, since our outcome is REM-bout duration rather than REM pressure. HREMSP has no established marker~\citep{ginsberg2024predictive}; proposed measures are either timing-based~\citep{benington1994rem, bassi2009time} or, as in our analysis, amount-based, where $\mathrm{REM_{post}}$ duration is treated as an index of accumulated propensity. This has been the working assumption under which probabilistic propensity measures have been validated in rodents~\citep{ginsberg2024predictive, park2021probabilistic}. Within these models, the REM hourglass is proposed to govern both transition probability and episode consolidation~\citep{ocampo2020rem}.

The stronger $\beta$ for $\mathrm{REM_{pre}}$--$N_s$ than for $N_s$--$\mathrm{REM_{post}}$ could also suggest an alternative interpretation: that an NREM sleep debt accumulates during REM sleep and is subsequently discharged during NREM sleep. Moreover, circadian and homeostatic components cannot be fully separated within one night, despite the cycle-by-cycle analysis. Forced-desynchrony protocols~\citep{wang2023desynchronizing} and more interventional models in suprachiasmatic-nucleus-lesioned animals would be needed to investigate this hypothesis further.

\begin{collaboration}
The operating hypothesis, that the bimodal distribution of inter-REM interval (IRI) durations reflects a robust architectural feature, emerged through human--AI collaboration. The investigator initially sought to extend an existing REM-homeostasis framework into a two-process-style model, while the agent's iterations surfaced a weak fit for that model alongside a highly robust IRI bimodality, redirecting the investigation towards the latter. The investigator selected the three target cohorts (BioSerenity, SSC, and CDH), fixed the primary covariate set (age, sex, BMI, AHI), and pre-specified the sensitivity scope. Within these constraints, \method designed and executed the bout-segmentation and statistical pipelines, performed the cross-cohort analyses and the NT1 application, generated all figures, and refined emerging sub-hypotheses across eight iterative rounds of investigator feedback. The agent was most responsive early in the process; broadly scoped instructions were occasionally misinterpreted or incompletely retained during later rounds. The principal benefit was compression of the exploratory phase: a relevant prompt, grounded in the field literature, could be prepared in less than an hour and dispatched overnight, with initial results available the following morning to determine whether the direction warranted further investigation.
\end{collaboration}

\section*{Transient oscillation disruption in narcolepsy type~1}

\begin{mainfinding}
A comprehensive scan of transient oscillations, defined as brief EEG bursts spanning 4--25\,Hz, identified two robust, multi-site signatures: a frontal fast-sigma (spindle-band) deficit during N2/N3 sleep, independently validated using YASA spindle density; and a coherent redistribution towards slower rhythms, with elevated theta during Wake/NREM and low-alpha during Wake, N1, and REM. Beyond these robust signatures, slow-oscillation power was reduced across Wake, N1, N2, and REM, while the reorganization of transient-oscillation coupling to slow-oscillation phase was more site-sensitive and is therefore considered exploratory.
\end{mainfinding}

\begin{figure}[!htbp]
\centering
  \includegraphics[width=\linewidth]{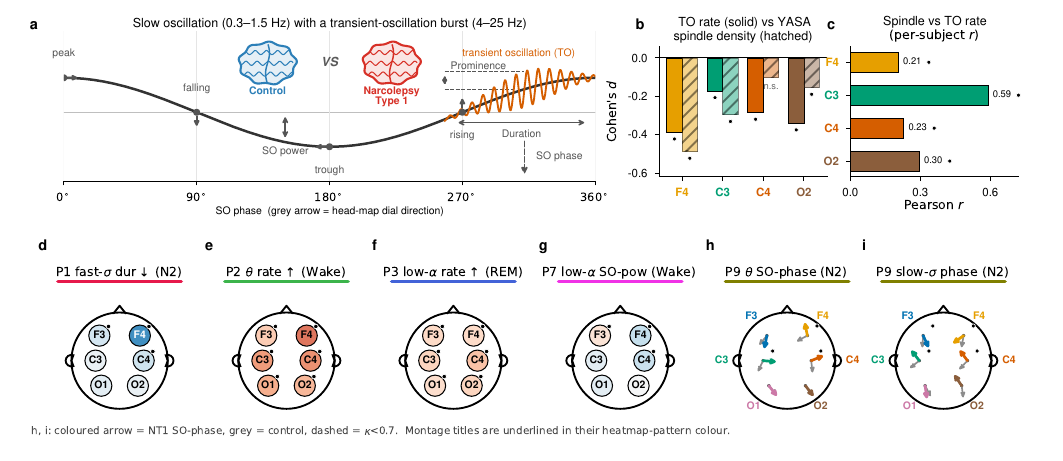}
  \caption{\textbf{Transient-oscillation signatures of narcolepsy type~1.}
  Throughout, head-maps in (d--g) are filled by effect size, Cohen's $d$
  (NT1 $-$ control): \emph{red} = higher in NT1, \emph{blue} = lower in NT1, white
  $\approx 0$, on the diverging scale shown in Supplementary~\Cref{fig:to_nt1_supp} ($-2$ to $+2$); a filled dot marks FDR $q<0.05$. In (h,i)
  colour instead encodes channel identity, not effect size.
  (a)~Anatomy of a transient oscillation (TO) riding the up-state of a slow
  oscillation (SO), defining the per-TO features: \emph{duration}, \emph{prominence},
  \emph{SO-power}, \emph{SO-phase}, and \emph{rate} (TO count per stage); the inset
  contrasts a control spindle with the reduced NT1 spindle (SO 0.3--1.5\,Hz,
  TO 4--25\,Hz). The SO-phase dial sets the head-map convention.
  (b)~Validation with an independent detector: Cohen's $d$ (NT1 versus control) for
  fast-$\sigma$ TO rate (solid) and YASA spindle density (hatched) at F4, C3, C4, O2.
  (c)~Per-subject Pearson correlation between fast-$\sigma$ TO rate and YASA spindle
  density.
  (d--g)~Topographic detail for representative direction-based patterns:
  (d)~P1 fast-$\sigma$ duration reduction (N2); (e)~P2 $\theta$ rate excess (Wake);
  (f)~P3 low-$\alpha$ rate increase (REM); (g)~P7 low-$\alpha$ SO-power (Wake). In (g),
  SO-power is reduced (blue) at all channels except F3, which shows a slight increase
  (red, $d=+0.28$); this left-frontal sparing is the defining feature of pattern P7.
  (h,i)~N2 SO-phase angle (P9) for $\theta$ and slow-$\sigma$: each head-map arrow gives
  the mean SO-phase (coloured = NT1, grey = control; dashed when reliability is low,
  $\kappa < 0.7$), read against the dial in (a); arrow \emph{colour} denotes the channel
  (F3, F4, C3, C4, O1, O2), not effect size.
  Panel titles are underlined in the colour of their pattern (P1--P9); the comprehensive
  feature-by-region scan from which these patterns are drawn, and the shared colour scale,
  are shown in Supplementary~\Cref{fig:to_nt1_supp}.}
  \label{fig:to_nt1}
\end{figure}

Conventional polysomnographic staging resolves sleep at the level of 30-second epochs, whereas the pathophysiology of narcolepsy type~1, which involves the selective loss of hypocretin (orexin) neurons, is expected to perturb sleep microstructure at much finer timescales. The biological question was whether a comprehensive scan of transient oscillations (TOs), brief, spectrally resolved EEG bursts detected using the Dynamic Oscillation Toolbox (DynamO)~\citep{stokes2023transient} across the full time--frequency plane (4--25\,Hz), would reveal coherent biological structure beyond the handful of spindle measures examined in prior single-site work, and whether such structure could be recovered even when individual effects fell below the significance threshold.

Transient oscillations are brief, localized bursts of oscillatory activity in the sleep EEG that appear as peaks in the time--frequency representation of the signal. They reflect transient increases in rhythmic neural activity at a characteristic frequency before dissipating. Sleep spindles (12--16\,Hz bursts lasting 0.5--2~s) are the best-known example, but TOs encompass a broader class of events spanning a wide range of frequencies (4--25\,Hz), including theta (4--8\,Hz), alpha (8--11\,Hz), sigma (9--16\,Hz), and beta (16--25\,Hz). The key insight is that traditional spindle detectors apply amplitude and duration criteria historically calibrated to events visible by eye and therefore select only a high-amplitude subset of a broader continuum of transient events~\citep{dimitrov2021,stokes2023transient}.

The study was carried out end to end in the CDH cohort, a six-site clinical PSG archive comprising 482 individuals with NT1 and 417 controls~\citep{stephansen2018neural,zhang2018national}. Six TO features (rate, prominence, duration, SO-phase coupling strength, SO-phase angle, and SO-power; \Cref{fig:to_nt1}a) were computed for each channel, sleep stage, and frequency band and residualized for age, sex, BMI, and site. All 900 features were then tested with Benjamini--Hochberg FDR correction (Supplementary~\Cref{fig:to_nt1_supp}). Two complementary strategies were applied: conventional FDR-based grouping and a direction-based coherent-pattern analysis that identifies feature--stage combinations with consistent effect directions across channels and frequency bands, even when individual effects are subthreshold. Expert feedback then guided the synthesis of these findings into five biologically interpretable themes (\Cref{fig:to_nt1}).

Two themes emerged as robust NT1-associated signatures. First, NREM sleep showed a fast-sigma deficiency: spindle-band (12--16\,Hz) TOs were less frequent and, with larger effect sizes, shorter and lower in amplitude during N2/N3, with the strongest effects observed frontally (F4 N2 rate: $d=-0.78$, 95\% CI $-1.03$ to $-0.50$, $q=4.8\times10^{-9}$; F4 N2 duration: $d=-1.23$, 95\% CI $-1.54$ to $-0.87$, $q=2.2\times10^{-15}$, the largest effect in the scan outside the circular SO-phase measures; $n=102$ NT1/191 controls; \Cref{fig:to_nt1}d). This deficiency was further validated using YASA spindle density (F4: $d=-0.49$, 95\% CI $-0.82$ to $-0.16$, $n=64/112$; C3: $d=-0.30$, 95\% CI $-0.48$ to $-0.11$, $n=223/284$; \Cref{fig:to_nt1}b,c).

Second, oscillatory activity was redistributed. Theta rate and duration increased during Wake and NREM sleep, with the largest effect observed at F4 during Wake (rate $d=+1.07$, 95\% CI $+0.77$ to $+1.35$, $q=2.7\times10^{-12}$; \Cref{fig:to_nt1}e). Low-alpha rate increased consistently across Wake, N1, and REM (18/18 features showed positive effects; mean $d=+0.35$; \Cref{fig:to_nt1}f), whereas beta rate and duration decreased (F4 Wake rate: $d=-0.80$, 95\% CI $-1.06$ to $-0.56$, $q=7.4\times10^{-9}$). Together, these findings indicate a coherent shift towards slower oscillatory modes, consistent with the state-boundary instability associated with hypocretin loss~\citep{saper2010sleep,itoyamanaka2023}.

The direction-based analysis assessed whether the sign of an effect was consistent across all six channels within a frequency band, regardless of whether individual features passed FDR correction. This analysis recovered coherent structure that significance thresholds alone missed, and that may warrant further investigation. The most consistent signal in the dataset was a cross-stage reduction in slow-oscillation (SO) power: every N1 and N2 feature showed the same effect direction (100\% directional consistency), as did the large majority of Wake and REM features (Supplementary~\Cref{fig:to_nt1_supp}), with left-frontal sparing during Wake and REM (\Cref{fig:to_nt1}g). A second emergent theme was coupling reorganization. During N2, the preferred SO phase of every non-theta band shifted coherently, with the largest effects observed centrally (low-alpha $d=1.42$--$1.49$; \Cref{fig:to_nt1}h,i), whereas theta alone broke this coherence. One possible interpretation, suggested by \method but not tested here, is that theta activity arises from hippocampal--cortical rather than thalamocortical generators.

Two additional patterns were considered exploratory. The first was a weak prominence asymmetry (P6; Supplementary~\Cref{fig:to_nt1_supp}): 129 of 150 prominence features showed a consistent left/right split, with greater prominence in NT1 than in controls at F3, C3, and O2 and lower prominence at F4, C4, and O1. The second comprised large N2 SO-phase-angle shifts (P9; \Cref{fig:to_nt1}h,i), which attenuated after stratification by site. Neither pattern is interpreted as an established finding.

Taken together, centrofrontal-channel fast-sigma reduction and theta excess, mostly detectable in N2, emerge as robust, multi-site NT1-associated signatures of sleep microstructure. The physiological significance of this shift from sigma to theta is uncertain; most PSG studies to date have focused on N1, Wake, and REM sleep, where such abnormalities are typically found~\citep{stephansen2018neural}. In contrast, spindle density has generally been reported as normal in NT1~\citep{cornelissen2020covalent}. The finding here suggests that abnormal distribution features beyond spindle density are present in NT1. Spindles are generally considered an electrophysiological biomarker of cognitive capacity and thalamocortical circuit integrity. Such a defect could thus partially explain cognitive difficulties beyond sleepiness that are reported by patients with NT1/orexin deficiency~\citep{lammers2026effects}. Both fast-sigma reduction and theta excess sit within a broader redistribution of oscillatory rate and diminished slow-oscillation background, while the SO-phase-coupling findings, which attenuate under site stratification, are reported as hypothesis-generating. This case illustrates the value of three elements in combination: a comprehensive scan, a direction-based analysis that treats consistent effect sign as evidence, and expert synthesis that both organizes the findings and tests them against confounds. It also revealed a new microarchitectural abnormality in NT1.

\begin{collaboration}
The investigator set the broad direction of a sub-epoch TO scan in the CDH cohort, and the agent proposed analytical strategies to support it. The investigator then fixed the design: the NT1-versus-control contrast, the six TO feature families (rate, prominence, duration, SO-phase coupling strength, SO-phase angle, SO-power), the covariate set (age, sex, BMI, site), and a dual FDR-corrected and direction-based analysis. Within that design, \method computed all features, ran both analyses, screened artefacts, performed the site-stratified confound checks, and generated the figures. The study was run as several parallel branches with deliberately different methodological choices, each refined over roughly five to eight rounds of feedback; only findings that reproduced across branches were reported. Feedback was mostly methodological: sensitivity analyses (e.g. site-stratification), prominence artefact removal (16 outliers removed), and scoping or figure edits. It was especially useful for probing \method's non-obvious choices. The earliest iterations needed the most attention; once the analytical contract stabilized, changes were usually integrated cleanly, and on long chains a fresh restart proved more effective than continued steering.
\end{collaboration}

\section*{Discussion}
\label{section_discussion}

We applied \method to five case studies in four clinical and epidemiological PSG cohorts, spanning disease risk, clinical phenotyping, and mechanism. Each case study recovered established features of sleep physiology before addressing a question that had not previously been examined at this scale: coupling between organ systems during sleep, the domain composition of physiological ageing, the temporal organization of cortical arousals, NREM--REM cycling structure, and the organization of transient oscillations in narcolepsy type~1. All five were refined through successive rounds of expert feedback, and selected findings were assessed through independent reruns, external validation, or cross-site replication where the available data and study design allowed. The evaluation therefore reflects repeated, expert-reviewed workflows rather than the output of a single unattended run.

In both disease-risk studies the disease signal was specific rather than global. In the coupling study, alterations were specific to disorder, sleep stage, and subnetwork: reduced N2 brain--brain coupling was associated with a higher risk of subsequent Parkinson's disease and Alzheimer's disease diagnoses, and reduced brain--heart coupling with subsequent Parkinson's disease, Alzheimer's disease, and dementia. Both are biologically plausible. Progressive loss of cardiac noradrenergic sympathetic innervation is characteristic of Parkinson's disease and can occur independently of nigrostriatal dopaminergic degeneration~\citep{candia2024framework}. One possible explanation for the brain--brain findings is that deficits in dopamine in Parkinson's disease and acetylcholine in Alzheimer's disease, together with proteinopathy-related synaptic and interneuron impairment, disrupt the excitatory--inhibitory balance and structural connectivity required for coordinated oscillatory coupling. Cross-site concordance of these effect-size patterns was only moderate. In the sleep-ageing study the specificity appeared at the level of organ systems instead. Decomposing PSG ageing into physiologically defined domains yielded a more interpretable representation that generalized better across cohorts than an unconstrained model. This suggests that future sleep-ageing models should preserve modular structure across neural, respiratory, autonomic, and motor systems, and should test adaptive or conditionally weighted fusion rather than simply increasing feature dimensionality. The resulting residual should not be interpreted as a universal biological-age axis: depending on the outcome, it may reflect shared age-related vulnerability, disease-specific pathophysiology, consequences of established disease, or combinations of these processes~\citep{lopezotin2023hallmarks}. Disease-tuned scores remain useful secondary risk models, but in this analysis they did not replace the age-derived late-fusion residual as the primary ageing construct.

Our findings refine current models of COMISA by suggesting that its distinguishing physiology lies less in an exaggerated burden of respiratory events than in an altered response to those events~\citep{hoc2025clinical, wulterkens2024heart, brooker2023obstructive}. The prolonged wakefulness and greater irregularity following cortical arousals support the hypothesis that impaired re-establishment of stable sleep, rather than increased respiratory instability itself, may be a defining feature of the disorder~\citep{wulterkens2024heart, brooker2023obstructive}. This interpretation is consistent with concepts of insomnia-related hyperarousal while also explaining why COMISA remained closely aligned with OSA across most sleep-microstructure measures~\citep{popovici2025hidden, wulterkens2024heart}. More broadly, these results highlight the value of moving beyond conventional PSG summary metrics towards dynamic measures of cortical arousal organization, which may better capture clinically relevant heterogeneity in sleep disorders~\citep{hanif2026use, Hanif2026_REM}. Although the observational design precludes causal inference and the degree of OSA skew varied across cohorts, the consistent prominence of post-arousal dynamics suggests these measures warrant investigation as markers for stratifying COMISA and for identifying patients who might benefit from interventions targeting both respiratory events and post-arousal sleep stabilization.

Within inter-REM intervals, subsequent REM-bout duration tracked intervening NREM duration more closely than intervening wake, reproducing a relationship previously reported in rodents and small human samples and holding under the disordered REM regulation of narcolepsy type~1. That both associations were significant is compatible with separate short-term and long-term components of REM homeostasis~\citep{franken2002long}, and because waking time over 24 hours exceeds NREM time roughly threefold, their net daily contributions could still be comparable. Inter-REM NREM duration was also bimodally distributed, with the valley closely preceding the first N3 epoch and coinciding with rising slow-wave density; this temporal association is compatible with, but does not establish, a transient inhibition of the NREM-to-REM transition that would favour slow-wave consolidation. The extent to which these associations inform REM homeostasis itself warrants caution. The outcome is REM-bout duration rather than REM pressure, for which no established marker exists~\citep{ginsberg2024predictive}; circadian and homeostatic contributions cannot be separated within a single night; and an alternative reading, in which NREM debt accumulates during REM sleep and is discharged during NREM sleep, is also compatible with the observed asymmetry. Forced-desynchrony protocols~\citep{wang2023desynchronizing} or interventional animal models would be needed to distinguish them. In narcolepsy type~1, the fast-sigma deficit and centrofrontal theta excess point to microarchitectural pathology that 30-second staging does not resolve. Spindle density has generally been reported as normal in this disorder~\citep{cornelissen2020covalent}, so the abnormality involves properties of spindle-band events beyond their density. Because spindles are considered a marker of cognitive capacity and thalamocortical circuit integrity, such a defect could partially explain the cognitive difficulties patients report beyond sleepiness~\citep{lammers2026effects}.

Obtaining these results depended on an expert-directed division of labour that shifted across the pipeline. At the hypothesis stage the most productive directions arose when experts seeded the system with relevant literature and scientific intuition, after which \method expanded the candidate space and assessed which questions the data could support. A focused, reference-seeded prompt could be dispatched overnight, letting an investigator survey the conceptual neighbourhood of an idea before committing to it. During preprocessing, pipelines were developed and tested on sample recordings, with expert review before full-cohort execution aligning methodological choices with domain conventions and checking the physiological plausibility of the resulting features. Execution was the most autonomous stage: once the analytical contract was fixed, the study proceeded through staged analysis, internal critique, and reporting, with the expert intervening at checkpoints rather than directing every step. Two design choices made this workable at scale. First, because the cohorts hold more than 50\,TB of raw signal, expensive preprocessing is separated from hypothesis-specific analysis and validated features are cached for reuse, amortizing the cost of raw-signal processing across studies. Second, because PSG spans systems with distinct noise characteristics and clinical conventions, cohort definitions, canonical splits, channel conventions, and analytical guidance are encoded in the shared substrate rather than restated in every prompt, which narrows the space of plausible but inappropriate analytical choices.

Several lessons emerged across the investigators who used the system, and each of them changed how it was built or used over the course of the case studies. Prompt structure mattered: broad or multi-part requests produced diffuse analyses, whereas narrow prompts with a single objective yielded more focused and reproducible outputs, and investigators therefore converged on decomposing complex aims into self-contained tasks. Long-horizon work degraded in a characteristic way, with the system drifting from the original question, carrying forward early assumptions, or letting an error introduced in one iteration shape later analyses as earlier constraints were revised or lost. Restarting a step with a consolidated specification proved more effective than appending further feedback, and became the standard response once an analytical contract had drifted. Neither transparency nor fidelity to the evidence was automatic. Early runs had to be asked explicitly for the cohort-flow tables, exclusion summaries, and intermediate artefacts needed to evaluate a result, and successive rounds of feedback and revision of the system made these outputs a routine part of every run. The provenance artefacts described in the Methods, including the cohort-flow record whose arithmetic is checked mechanically, address the same gap. Contributions also ran in both directions: within an expert-defined contract the system proposed useful methodological changes, including the age-bias correction adopted in the sleep-ageing study, functioning less as an execution tool than as a collaborator whose suggestions could be evaluated and revised.

There are several limitations in this work. The analyses are retrospective and rest on clinical-referral or epidemiological cohorts rather than general-population samples, making them subject to the ascertainment biases inherent in how these recordings were collected. The reported associations should be treated as hypothesis-generating rather than causal, and require designs that emulate a target trial, and ultimately prospective or interventional studies, for confirmation. They also rest on single-night, in-laboratory PSG, which is subject to the first-night effect and does not capture night-to-night variability at home; a single recording is an imperfect estimate of habitual sleep, and longitudinal designs are needed to separate state from trait~\citep{agnew1966first}. The case studies vary in their proximity to clinical translation: the COMISA arousal-recovery phenotype and the sleep-age residuals rely on routinely derived PSG features, whereas network-physiology coupling and transient-oscillation analysis require full-bandwidth multichannel recordings and specialized post-processing that most clinical systems do not support. None has been evaluated as a clinical decision tool, and actionability will require prospective studies showing that acting on these measures improves outcomes. The scope is bounded in other ways: the hypothesis and execution stages operated mainly on precomputed features rather than raw signals, and five case studies sample only a small region of a vast hypothesis space. Finally, automated critique can catch procedural and internal-consistency errors but is not a dependable arbiter of scientific novelty or importance, so judging whether an open-ended finding is genuinely new still rests with domain experts. The safeguards reduce but do not eliminate hallucination, and subtle fabricated or misattributed content can survive automated checks, particularly as errors accumulate across revisions. Behaviour also varied across model versions and capability tiers, which presents a reproducibility challenge as deployed models change.

Several directions could extend this work. Larger and more representative cohorts, including general-population and prospective cohorts with multi-night recordings, would help separate trait from state, and samples collected under protocols that dissociate homeostatic from circadian influences, such as constant-routine and forced-desynchrony designs, would allow the mechanistic questions raised here to be tested directly. Additional patient cohorts are also needed, as some pathologies uniquely challenge existing assumptions. A next generation of the system could complement its curated feature library with direct access to raw signals and learned representations, including SleepFM embeddings~\citep{thapa2024sleepfm}, enabling patterns not captured by predefined features to be discovered and interrogated. Ultimately, cross-modal retrieval across PSG, electronic health records, imaging, wearables, and genomics could support hypotheses spanning multiple levels of physiology. The deeper challenge is to widen the portion of the scientific process that can be automated without sacrificing rigour, where the bottleneck is evaluation rather than generation. Open problems include a rigorous automated judge of methodological soundness and claim calibration, a validator that independently reproduces and stress-tests a result rather than confirming it as written, and reliable measures of genuine hypothesis novelty. One resource for that effort may be a by-product of the workflow itself: because every run preserves the collaboration trajectory behind a study, including the questions posed, the corrections made, and the judgements that shaped the analysis, these records could in principle train future models to act as more effective scientific collaborators.

\section*{Acknowledgements}

The Human Sleep Project is supported by a grant from the National Heart, Lung, and Blood Institute of the NIH (R01HL161253).

R.T. is supported by Knight-Hennessy Scholars funding. E.M. is supported by a grant from the National Heart, Lung, and Blood Institute of the NIH (R01HL161253). J.Z. is supported by funding from the Chan-Zuckerberg Biohub.
We thank Topi Niemi and Esa Räsänen for their help with the preprocessing steps to calculate the HRV features.

\section*{Author Contributions Statement}

R.T., E.M., and J.Z. conceived the project. R.T. led the study, developed the agent system, and coordinated the five case studies. R.T. and M.R.K. built the initial iteration of the agent infrastructure. U.H., R.G., A.S., A.B.-K., and E.H. preprocessed the expert-led datasets. R.G. led the REM-sleep regulation case study, A.S. led the transient-oscillation study, A.B.-K. led the sleep-ageing study, U.H. led the COMISA study, and M.S. led the network-physiology study. H.G.Z. provided clinical expertise and guidance on agentic system design. E.C.L. provided domain-expert feedback that guided the iterative refinement of AI-generated hypotheses, analyses, and scientific interpretation. E.M. and J.Z. supervised the project. All authors contributed to writing and copyediting the manuscript.



\clearpage
\newpage

\bibliographystyle{unsrt}
\bibliography{references}

\section*{Methods}
\label{section_methods}

\subsection*{Cohorts and data}
The shared data substrate comprised four clinical and epidemiological PSG cohorts (\Cref{tab:cohort_demographics}). Raw signals were stored in a harmonized 128\,Hz HDF5 representation, with channels retaining cohort- and site-specific names. A shared channel-group registry mapped physiological modalities to the corresponding channel names at each site, allowing downstream pipelines to operate across cohorts without hard-coded site-specific channel labels. Each case study applied its own eligibility, feature-availability, and quality-control criteria; the resulting analytical samples are therefore reported separately in the corresponding cohort-flow tables.

\paragraph{SSC.}
The Stanford Sleep Clinic (SSC)~\citep{thapa2024sleepfm,kjaer2025stanford} is a clinical PSG cohort acquired at Stanford University Medical Center. Linked electronic health records provide demographics, ICD-coded diagnoses, medications, BMI, and longitudinal follow-up for incident disease analyses.

\paragraph{HSP.}
The Human Sleep Project (HSP v2.0)~\citep{li2026hsp} is a clinical PSG cohort available through the BIDMC Brain Data Science Portal (\url{https://bdsp.io/content/hsp/2.0}). Linked electronic health records provide demographics, diagnoses, and longitudinal diagnostic follow-up used for external validation.

\paragraph{BioSerenity.}
The BioSerenity clinical network~\citep{hanif2023automatic,hanif2024associations,hanif2026use} comprises PSG recordings collected across a multi-site US clinical setting. Available information includes respiratory measures, sleep questionnaires, comorbidity history, and self-reported sleep symptoms. The PSG substrate is distinct from the larger questionnaire database used as the starting source for the COMISA cohort flow.

\paragraph{CDH.}
The Central Disorders of Hypersomnolence (CDH)~\citep{stephansen2018neural,zhang2018national} registry is a Stanford-curated multi-site cohort assembled from seven institutions across France, China, Italy, Austria, Korea, and Denmark. It includes harmonized diagnoses of narcolepsy type~1, narcolepsy type~2 (NT2), idiopathic hypersomnia, secondary hypersomnia, and matched controls.

\subsection*{Software and reproducibility}

\method was implemented in Python~3.11 using the Claude Agent SDK (v0.1.76) and the Claude Code command-line runtime (\texttt{@anthropic-ai/claude-code}). All language-model calls used Claude Opus~4.6 (\texttt{claude-opus-4-6}). Agents executed inside a Singularity container built from a version-controlled definition file (\texttt{python\_node\_3.11\_22.def}) containing Python~3.11, Node~22, the statistical analysis stack (\texttt{lifelines}, \texttt{statsmodels}, \texttt{scikit-learn}, \texttt{numpy}, \texttt{scipy}, and \texttt{mne}), and the version-pinned feature-extraction tools required by the relevant pipelines (Supplementary~\Cref{tab:features}). Jobs were submitted to a shared HPC cluster using the same container environment for development, sample-scale validation, and full-cohort processing. Each run generated a versioned workspace containing the executed scripts, tool logs, sub-agent prompts, iteration summaries, structured critique verdicts, and resulting analysis artefacts. These workspaces preserve the analytical trajectory and provide the primary reproducibility artefacts for each case study. Because language-model generation is stochastic, independently rerunning an agent is not expected to reproduce the same trajectory exactly; instead, the reported numerical results are reproduced from the preserved analysis scripts and fixed data inputs.

\subsection*{System architecture}

\method comprises an Interface Layer and three specialist agents that can be invoked individually or iteratively as a research question develops. The Interface Layer provides the conversational surface through which the expert directs work, launches agents, reviews outputs, and supplies feedback. The Hypothesis Agent converts a research direction into a scored portfolio of data-grounded candidate questions. The Preprocessing Agent develops and validates feature-extraction pipelines from raw PSG signals on sample recordings. The Execution Agent carries a selected hypothesis through statistical analysis to a draft manuscript with code-grounded results. The three specialist agents are launched and reviewed through the Interface Layer, which maintains the persistent conversational state for each project; their analytical artefacts and intermediate outputs persist in versioned workspaces.

Three design commitments span these components. First, filesystem-based continuity: successive specialist-agent launches do not depend on persistent model context, but exchange state through versioned files on shared storage. This makes interrupted or iterative runs resumable and preserves an audit trail of prompts, tool calls, outputs, and feedback. Second, code-grounding: each quantitative result reported by the Execution Agent is linked to the executed analysis script, input data product, and output artefact from which it was derived. Third, expert-controlled checkpoints: in the expert-guided configuration evaluated here, the agents autonomously develop sample-scale preprocessing pipelines, execute analyses, and draft reports, while the expert approves full-cohort feature extraction, selects hypotheses for execution, and determines whether a manuscript is ready for finalization.

\subsubsection*{Shared infrastructure}

All four components operate against a shared substrate that defines which data are available and how they can be accessed. Preprocessed feature sources are exposed through a uniform \emph{describe}, \emph{list}, and \emph{load} interface implemented as a Model Context Protocol (MCP) server. Agents therefore inspect and access available data through schema-validated calls rather than relying on prompt-embedded file documentation that may become inconsistent with the underlying data. A canonical data-split function provides the authoritative top-level assignment of recordings to training, validation, and test sets, where applicable. Downstream analyses may construct folds within the training partition but cannot redefine validation or test membership. This keeps reported performance comparable across runs and guards against data leakage, which an agent free to define its own partition could otherwise introduce. The split definition is mounted separately from the selected cohort, and the container does not launch if this mount is absent. A shared skills library provides reusable procedural guidance and enforced figure-style standards, including colour-blind-safe palettes, typography, and axis-labelling conventions, thereby providing a common source of visual and methodological guidance across agents.

Each specialist agent runs inside a Singularity container with deliberately asymmetric filesystem permissions. System code, shared libraries, cohort features, raw signals, the hypothesis specification, and expert-feedback files are mounted read-only. The per-experiment workspace is writable and contains the analysis code, intermediate results, figures, reports, and logs generated during the run. These permissions are enforced by the container rather than by prompt instructions, preventing an agent from modifying source cohort data, authoritative split assignments, or its governing research specification. Session logs and machine-readable records of tool calls, including their inputs, outputs, and issuing agent, are written automatically to the experiment workspace, allowing actions by lead and sub-agents to be distinguished during post-hoc audit.

\subsubsection*{Interface Layer}

The Interface Layer is a FastAPI web application through which the expert directs the system, reviews outputs, and provides feedback. It is the only component that maintains persistent conversational state for each project across sessions; specialist-agent state is instead preserved through the versioned workspace artefacts described above.

The interface serves three roles. First, it acts as a thinking partner during question formulation, helping the expert relate the scientific objective to the available cohorts and determine what evidence would constitute a convincing result. These discussions are grounded in the available data through the shared MCP interface. Second, it launches and monitors specialist agents through an MCP orchestration server that manages project lifecycle, hypothesis curation, Slurm job submission, status polling, and feedback authoring. Server-side validation restricts launches to supported cohorts and launch modes; in the expert-guided workflow evaluated here, launch parameters are displayed for confirmation before a cluster job is submitted. Third, it provides an audit and feedback channel. The interface reads specialist-agent artefacts (critique verdicts, cohort-flow records, iteration summaries, figures, and manuscript drafts) and converts expert comments into numbered, file-anchored actions while preserving the original feedback verbatim. It then supplies this feedback file to the relevant specialist agent, which treats it as the highest-priority input during the subsequent iteration.

The Interface Layer maintains one persistent Claude Agent SDK session per project. The session is reused across chat turns so that prior messages, file reads, and tool calls remain available without being reintroduced manually. When a project session is initialized, its system prompt is assembled from a static role specification, project metadata, and a project-memory file (\texttt{MEMORY.md}). The interface updates this file periodically as a concise record of project status and outstanding decisions rather than as a complete transcript. The SDK session store preserves conversational continuity across server restarts by replaying the stored session, while the project workspace preserves the corresponding analytical artefacts.

\subsubsection*{Hypothesis Agent}

Given a research direction and one or more target cohorts, the Hypothesis Agent produces a scored portfolio of candidate scientific questions while preserving the lineage of each candidate across refinement steps. It runs parallel refinement trajectories within predefined diversity bins spanning scientific domains and eligible cohort contexts. At each step, several variants are generated from the current candidate and the highest-scoring variant is retained for further refinement. Four sub-agents operate under a lead orchestrator: a knowledge aggregator that assembles relevant literature and prior hypotheses; a generator that proposes candidate questions using this context and web-based literature retrieval; a critic that evaluates candidates against a multidimensional rubric; and a portfolio manager that records selected and rejected candidates, maintains the search graph, and assembles the final portfolio.

Two gates are applied before a candidate enters the search graph. The first assesses data feasibility. Variables named in a hypothesis are checked through the shared MCP interface against the cohort data catalogue, including preprocessed physiological features, demographics, clinical metadata, and available outcomes. If a required physiological feature is unavailable, the raw-signal channel registry is queried to determine whether it could be derived through a new preprocessing pipeline. Candidates requiring variables that are neither available nor derivable are rejected before scoring. The second gate audits the candidate's scientific-question contract. Agent-generated candidates that assert anticipated effect sizes or sample sizes, or unnecessarily prescribe statistical methods and covariates, receive a capped score so that unsupported analytical detail is not mistaken for scientific specificity. This restriction applies to candidate generation; an expert may subsequently provide prespecified methods, covariates, or other requirements as part of the execution contract.

Each candidate is independently evaluated by the critic at two sampling temperatures. Disagreement between the resulting scores reduces that candidate's priority during parent selection, limiting the influence of unstable evaluations. Search continues until the predefined round budget is exhausted or the trajectories meet stopping criteria for score saturation or loss of portfolio diversity. The resulting high-scoring, diverse portfolio is presented through the Interface Layer. In the expert-guided studies reported here, candidates were selected for execution by the expert rather than promoted automatically.

\subsubsection*{Preprocessing Agent}

When a hypothesis requires a physiological feature that is not already available in the shared data substrate, the Preprocessing Agent develops and validates the corresponding extraction pipeline. Given a free-text feature specification and one or more target cohorts, it implements a generalizable pipeline and evaluates it on a small sample of recordings before any full-cohort processing is initiated. Its initial deliverables comprise sample-level feature outputs, a rerunnable extraction script, methods documentation, and an expert-handoff report describing the changes required to expose the feature to downstream agents. Full-cohort feature extraction is initiated only after expert review and approval.

The Preprocessing Agent comprises a lead orchestrator and four sub-agents: a researcher responsible for literature and tool discovery and raw-signal inspection; an executor that implements and tests the pipeline; a reporter that prepares the expert-handoff documentation; and a critic that reviews correctness, physiological validity, and generalizability. The lead coordinates the workflow and routes tasks and artefacts among these specialist roles.

Five principles govern preprocessing. First, execution is restricted initially to the sample defined in the feature specification, typically tens of recordings. Pipelines are parameterized so that the same extraction code can subsequently be applied to the full cohort by changing the sample restriction after expert approval. This sample-first design allows methodological and computational problems to be identified before committing substantial cluster resources. Second, pipelines cannot hard-code channel names. Each script queries the shared channel-group registry at runtime and intersects the registered modality mappings with the channels present in each recording, allowing site-specific naming differences to be handled without changes to the extraction logic. Third, the researcher evaluates established community tools before proposing a new implementation. When an appropriate published tool is available, reuse provides a validated methodological starting point, although its behaviour is still evaluated on the target cohort. Fourth, pipelines parallelize processing across recordings to support full-cohort execution on the HPC cluster. Fifth, the critic reviews each iteration for hard-coded channels, cohort-specific assumptions, missing-signal handling, and other implementation choices that could prevent generalization; unresolved violations are classified as critical issues.

In the final iteration, the reporter produces an expert-handoff document containing, for each proposed feature, the tool and version used, input channels, operational definition, units, output schema, sample-level distributions, quality-control results, and known limitations. It also includes validation figures and a checklist of properties requiring expert confirmation before full-cohort extraction. After approval and full-cohort processing, the resulting feature source is registered in the shared data-access library and exposed to downstream agents through the same MCP interface. The complete library therefore combines investigator-provided features and pipelines with features added through the Preprocessing Agent, as summarized in Supplementary~\Cref{tab:features}.

\subsubsection*{Execution Agent}

The Execution Agent carries a selected hypothesis through staged analysis to a draft manuscript with code-grounded results. Each quantitative claim included in the manuscript must be supported by a stored result artefact and linked to the Python script that generated it. The scripts, inputs, outputs, figures, and tables are retained in the experiment workspace for audit and, subject to data-access requirements, independent re-execution. The agent comprises a lead orchestrator and four sub-agents: a researcher responsible for literature review and dataset inspection; an executor that writes and runs the stage-specific analysis scripts; a reporter that drafts the manuscript during the final iteration; and a critic that independently reviews the analytical record and returns a structured verdict. Each sub-agent has a role-specific tool allowlist. In particular, the critic can inspect the project artefacts but cannot execute shell commands or modify files, structurally separating review from the generation of code and results.

\paragraph{Hypothesis and execution specification as contract.}
A central principle of the Execution Agent is that the selected hypothesis and its accompanying expert-provided specification jointly define the minimum required analysis. Concrete requirements, including the cohort, sample construction, model, covariates, outputs, and literature to consider, must be followed unless they are infeasible or the expert explicitly approves a change. The agent may add sensitivity analyses, alternative models, or robustness checks, but cannot silently omit or substitute a requested analysis. This prevents, for example, replacing a specified method with one judged to be ``approximately equivalent'' or analysing a smaller cohort because it is more convenient. Three mechanisms enforce this contract. First, required instructions are passed verbatim to each relevant sub-agent rather than only through an orchestrator's paraphrase, reducing the risk that methodological constraints are lost during delegation. Second, when the agent identifies a methodological concern, it retains the specified analysis when it remains statistically and computationally defined, clearly reports the concern, and performs the proposed alternative as an additional analysis for expert adjudication. Third, the agent halts or defers a required step only when it cannot be executed with the available data, tools, or computing environment, for example because a required file, variable, or software dependency is unavailable. In such cases, it records the diagnostic evidence and reports the unresolved requirement rather than substituting a different procedure.

\paragraph{Code-grounding.}
The Execution Agent requires each analysis workspace to contain a standard set of structured provenance artefacts. A fixed directory structure separates scripts, results, figures, and logs. Each analytical stage has a corresponding script with a machine-readable header documenting its purpose, inputs, outputs, parameters, and runtime. Before figure generation, a figures-and-tables plan maps each intended output to the script and data products that will generate it; subsequent changes to the plan are retained in the workspace history. A features ledger records each covariate and engineered variable together with its operational definition, units, plausible range where defined, and supporting literature where relevant. A methods ledger records analytical decisions, including their rationale, parameters, assumptions assessed, and sample sizes, which the reporter uses when drafting the Methods section. These ledgers reduce the risk that undocumented analytical choices are later described as prespecified or validated. A cohort-flow artefact records the starting population, sequential exclusions and their rationales, and the final analytical sample. Its arithmetic is checked mechanically, and it is rendered as the first cohort table in the internal case-study report. Across iterations, the critic compares successive cohort-flow versions and flags unexplained changes in sample construction. During the final iteration, the reporter creates a unified reproduction script that invokes the staged analysis code to regenerate the reported numerical results, figures, and tables from the fixed analytical inputs. This script provides a single entry point for reproducing the manuscript outputs; full raw-signal feature extraction remains a separate preprocessing step when rerunning it would be computationally prohibitive.

\paragraph{Automated quality control and hallucination mitigation.}
Beyond code-grounding, the Execution Agent applies automated checks for several common failure modes of language-model-generated scientific reports. Each citation is entered into a structured registry containing its title and source URL, allowing the reference itself to be traced to the retrieval event from which it originated. References whose bibliographic identity cannot be verified are flagged and excluded from the final agent-generated manuscript rather than completed from model memory. This check addresses fabricated or incorrectly specified references but does not, by itself, establish that a cited source supports the associated interpretation. Figure-style compliance is assessed in both the plotting code and the rendered output. Plotting scripts are checked for use of the shared figure-style package, while rendered PDFs are inspected for violations of the required palette, typography, labelling, and layout conventions. After manuscript drafting, standalone-readability checks flag workspace-specific paths, internal variable or feature names, and undefined acronyms that would prevent the report from being interpreted independently of its analysis workspace. These automated checks reduce identifiable errors but do not establish scientific validity, which remains subject to expert review.

\paragraph{Iteration model and expert feedback.}
At the end of each iteration, the critic returns a structured verdict comprising an overall quality label and counts of critical, important, and optional issues. When expert feedback has been supplied, any feedback item that remains unresolved is classified as critical and limits the overall verdict to ``Needs Improvement''. The orchestrator terminates the refinement loop when the critic reports no critical issues, when the configured iteration budget is exhausted, or when two consecutive iterations satisfy the predefined stagnation criterion. When the Interface Layer relaunches the Execution Agent with expert feedback, the lead creates an itemized response file that preserves each comment verbatim and records the corresponding required action, implementation status, and affected code, result, figure, or manuscript location. Relevant executor prompts include both the original comment and its translated action, reducing the risk that constraints are lost through paraphrase. Before invoking the critic, the lead marks each item as completed or deferred and provides supporting evidence or a rationale. The critic then independently inspects the cited artefacts and verifies whether each requested change has been implemented. Deferred items remain unresolved, and therefore critical, unless the expert explicitly accepts the deferral. Feedback iterations are non-destructive. Before a new iteration begins, the previous report, execution artefacts, and ledgers are copied to numbered version snapshots, allowing successive analyses and drafts to be compared. At the end of each iteration, the lead writes a structured summary of the completed work, unresolved issues, and current artefact locations. These summaries provide the primary filesystem-based memory for refinement across specialist-agent launches without requiring the complete prior conversation to be reloaded.

\subsection*{Data availability}

HSP v2.0 is distributed through the BIDMC Brain Data Science Portal (\url{https://bdsp.io/content/hsp/2.0}). The CDH cohort is available at \url{https://stanfordmedicine.app.box.com/s/r9e92ygq0erf7hn5re6j51aaggf50jly}, and SSC is available through the Brain Data Science Portal at \url{https://bdsp.io/content/08vg8vqv2wdtwonc1ddy/1.0/}. Access to linked electronic health records, longitudinal outcomes, and other restricted clinical variables is governed by cohort-specific approvals and may differ from the data included in the public PSG releases.

The BioSerenity data are proprietary and cannot be redistributed publicly. Requests for access are subject to approval by BioSerenity and the applicable institutional, privacy, and data-use requirements.



\begin{appendices}
\section{Supplementary Material}
\label{section_appendix}

\subsection{Preprocessed feature library (extended)}
\label{supp_features}

The shared feature library exposes preprocessed representations of raw PSG signals and associated metadata through a uniform Model Context Protocol (MCP) interface (see Methods). Feature availability across the four cohorts is summarized in Supplementary~\Cref{tab:features}. Features are grouped by pipeline origin. \emph{Human-led} denotes features or pipelines selected, supplied, or previously established by investigators, including curated clinical metadata and outputs from existing preprocessing systems. \emph{Agent-led} denotes pipelines curated, adapted, or implemented by the Preprocessing Agent under expert review; it does not imply that the underlying physiological method or software was invented by the agent. Feature definitions, output formats, and relevant cohort-specific considerations are described below.

\subsubsection{Human-led features}

\paragraph{Sleep stages (U-Sleep).}
U-Sleep~\citep{perslev2021u} is a fully convolutional neural network that maps EEG and EOG signals to sleep-stage probabilities. Outputs are stored as \emph{hypnodensities}: a $T \times 5$ probability matrix over Wake, N1, N2, N3, and REM at 1-second resolution. Hard-stage labels are derived by taking the highest-probability stage when required. Outputs are stored separately for each recording and available channel configuration, together with processing logs.

\paragraph{Cortical arousals (MAD).}
The Multimodal Arousal Detector (MAD)~\citep{brink2020automatic} identifies cortical arousals using available EEG, EOG, EMG, and ECG signals. Event-level outputs contain the onset, offset, duration, and predicted probability of each detected arousal.

\paragraph{Respiratory events (ABED).}
The Apnoea-Based Event Detector (ABED)~\citep{kjaer2026expert} identifies obstructive and central apnoea and hypopnoea events from available respiratory-effort, airflow, and oximetry signals as well as MAD-based arousal detections. Event-level outputs contain onset and offset times, event type, and either the predicted probability or corresponding binary classification.

\paragraph{EEG spectrograms.}
Time--frequency spectrograms were precomputed from available EEG channels to avoid repeating spectral decomposition during downstream analyses. Spectrograms were generated using the LSP-optimal multitaper spectrogram with 30-second non-overlapping windows. Spectral power was converted to decibel units and only frequencies between 0.5--25\,Hz were retained. Spectrograms are stored separately by recording and channel.

\paragraph{Transient oscillations (DynamO).}
The Dynamic Oscillation Toolbox (DynamO)~\citep{stokes2023transient} detects transient EEG oscillations, defined as brief, spectrally resolved bursts distributed across the time--frequency plane. For each channel, sleep stage, and frequency band, the stored outputs support six derived feature families: oscillation rate, prominence, duration, slow-oscillation phase-coupling strength, preferred slow-oscillation phase angle, and local slow-oscillation power.

\paragraph{Brain--heart coupling.}
Coupling between EEG and cardiac activity was computed for each available EEG channel using an established brain--heart interaction formulation~\citep{Saibene2026BrainHeart}. Outputs are stored as per-recording, per-channel arrays for downstream feature summarization.

\paragraph{Demographics.}
Available subject-level metadata include age, sex, BMI, race, and ethnicity. Availability and coding differ across cohorts, and race/ethnicity follows each cohort's native conventions (\Cref{tab:cohort_demographics}).

\paragraph{Diagnoses and curated outcomes.}
SSC and HSP contain linked ICD-9/ICD-10 codes and longitudinal time-to-diagnosis information. CDH contains harmonized primary and secondary diagnostic labels, including NT1, NT2, idiopathic hypersomnia, secondary hypersomnia, and controls. BioSerenity does not contain linked diagnostic codes in the shared substrate; the COMISA groups were instead derived from self-reported insomnia-related questionnaire and objective respiratory measures, as described in the corresponding case-study methods.

\paragraph{Medications.}
SSC contains longitudinal medication records extracted from linked electronic health records. Other cohorts either do not provide medication information or contain only limited, non-harmonized indicators; these were not exposed as a common medication feature source.

\paragraph{Sleep questionnaires.}
BioSerenity contains subject-reported sleep symptoms and behaviours, including the Epworth Sleepiness Scale and its component items, self-reported medical and problem history, and sleep-related symptoms and behaviours.

\subsubsection{Agent-led features}

\paragraph{R-peaks.}
R-peak locations were extracted from available ECG channels and represented as event times within each recording. These detections provide the input for downstream heart-rate-variability, brain--heart-coupling, and time-delay-stability analyses.

\paragraph{Spindles (YASA).}
Sleep spindles were detected during NREM sleep using YASA~\citep{vallat2021open}. Event-level outputs include spindle onset, duration, peak frequency, amplitude, and the EEG channel from which the event was detected.

\paragraph{Slow waves (YASA).}
Slow waves were detected during NREM sleep using YASA~\citep{vallat2021open}. Event-level outputs include onset, negative- and positive-peak timing, amplitude, duration, and the EEG channel from which the event was detected.

\paragraph{Eye movements (YASA).}
Eye-movement events were detected from available EOG channels using YASA~\citep{vallat2021open}. Event-level outputs include onset, offset, duration, detection channel, and event classification where available.

\paragraph{EEG artefact masks (YASA).}
YASA~\citep{vallat2021open} was used to identify EEG segments likely to contain artefact. The resulting time-aligned Boolean masks allow downstream pipelines to exclude contaminated periods from feature extraction and statistical summaries.

\paragraph{Continuous oximetry.}
Available pulse-oximetry signals were extracted into a standardized, time-aligned representation retaining the SpO$_2$ value and sampling information for each recording.

\paragraph{Desaturation events.}
Discrete oxygen-desaturation events were derived from the continuous SpO$_2$ signal. Event-level outputs include onset, offset, nadir, and the magnitude of the decrease from the preceding baseline.

\paragraph{Heart-rate variability.}
Heart-rate-variability features were calculated from the R-peak time series and stratified by sleep stage. The stored summaries include time-domain measures such as SDNN, RMSSD, and pNN50 and frequency-domain measures such as low-frequency power, high-frequency power, and their ratio.

\paragraph{Brain--EMG coupling.}
The brain--heart coupling framework was adapted to quantify stage-stratified coupling between EEG and available EMG signals. Outputs summarize cortical--muscular coupling for each available EEG--EMG channel pair.

\paragraph{Time-delay-stability networks.}
Time-delay stability (TDS)~\citep{bashan2012network} quantifies coupling among available cortical, cardiac, and peripheral physiological signals. For each signal pair, sliding-window cross-correlation was used to estimate the lag at which coupling was maximal. TDS link strength was defined as the proportion of windows in which this lag remained stable. A signal pair was classified as a significant link when more than 7\% of its windows met the stability criterion. Features were summarized by sleep stage and physiological subnetwork, including brain--brain, brain--heart, brain--body, and body--body coupling.

\subsection{Additional Case Study Details}
\label{supp:additional_case_study_details}

\subsubsection{Transient-oscillation features}
\label{supp:transient_oscillation}

Transient oscillations (TOs) were detected with the Dynamic Oscillation Toolbox (DynamO)~\citep{stokes2023transient}, which identifies them as \emph{time-frequency peaks} (TF-peaks):
salient, topographically distinct high-power regions in the multitaper spectrogram of the
EEG (1\,s windows, 0.05\,s steps, time-half-bandwidth 2, 3 tapers). A watershed
segmentation is applied to the negative of this topography; the resulting over-segmented
regions are merged under a completeness/distinctness weight rule, each surviving region is
trimmed to the boundary enclosing 80\% of its volume, and peaks falling below the
estimator's own time (0.5\,s) and frequency (2\,Hz) resolution are discarded as noise. We
used the toolbox's recommended \texttt{fast} preset: 30\,s segments, $2\times2$
topographic downsampling, and a merge threshold of 11, applied iteratively until no pair
of regions exceeded it. Because peaks below 4\,Hz cannot be disambiguated from the slow
oscillation itself, analysis was restricted to 4--25\,Hz and partitioned into five bands
($\theta$ 4--8, low-$\alpha$ 8--11, slow-$\sigma$ 9--12, fast-$\sigma$ 12--16, $\beta$
16--25\,Hz).

Six features summarize the TOs within each channel $\times$ stage $\times$ band:
\begin{itemize}
  \item \textbf{Rate}: number of TOs per unit time spent in that stage.
  \item \textbf{Log-prominence}: log of the TF-peak height above its local spectral
        baseline ($\mu$V$^2$/Hz), averaged over TOs; an amplitude measure.
  \item \textbf{Duration}: temporal width of the TF-peak (s), averaged over TOs.
  \item \textbf{SO-power}: slow-oscillation power (0.3--1.5\,Hz, integrated and
        expressed in dB from a separate 30\,s multitaper spectrogram) \emph{at the time
        of each TO}, averaged over TOs. This indexes the slow-wave background in which
        the TOs occur, not stage-wide SO power.
  \item \textbf{SO-phase angle}: circular mean of the slow-oscillation phase at TO
        times. Phase is obtained by band-passing the EEG at 0.3--1.5\,Hz with a
        zero-phase filter and taking the Hilbert angle, wrapped so that $0$\,rad is the
        SO peak and $\pm\pi$ the SO trough (surface-negative convention;
        \Cref{fig:to_nt1}a).
  \item \textbf{SO-phase coupling strength (MRL)}: mean resultant length of those same
        phases (0 = uniform across the SO cycle, 1 = perfectly phase-locked); it measures
        the \emph{concentration} of coupling, independently of the preferred angle.
\end{itemize}

\begin{figure}[!htbp]
\centering
  \includegraphics[width=\linewidth]{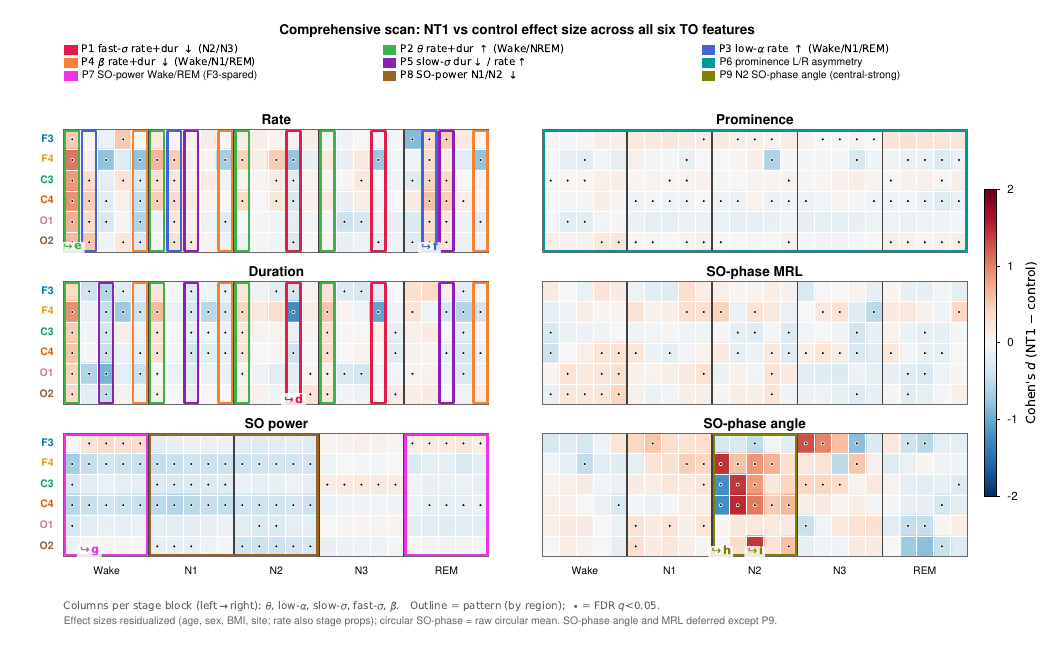}
  \caption{\textbf{Comprehensive transient-oscillation scan in NT1.}
  Cohen's $d$ (NT1$-$control, residualized for age, sex, BMI, and site) for all
  six TO features (rate, prominence, duration, SO-phase coupling strength,
  SO-phase angle, SO-power); each facet shows six channels (rows) $\times$ five stages $\times$ five
  frequency bands (columns; band order $\theta$, low-$\alpha$, slow-$\sigma$,
  fast-$\sigma$, $\beta$). Dots mark FDR $q < 0.05$; coloured outlines mark the
  nine direction-based patterns P1--P9 (legend), each defined by a consistent
  effect sign within a band/stage scope. Hooked arrows ($\hookrightarrow$d--i) on
  the bottom row of the relevant facets flag the feature shown in topographic detail
  in the corresponding panel of~\Cref{fig:to_nt1}. Circular SO-phase is the
  raw circular mean (not OLS-residualizable).}
  \label{fig:to_nt1_supp}
\end{figure}

Rate, log-prominence, duration and SO-power were residualized against age, sex, BMI and
site (rate additionally against stage proportions); the two circular measures are not
amenable to ordinary least-squares residualization and are reported as raw circular
statistics. The scan therefore comprises 6 features $\times$ 6 EEG channels (F3, F4, C3,
C4, O1, O2) $\times$ 5 sleep stages (Wake, N1, N2, N3, REM) $\times$ 5 frequency bands
$=$ \textbf{900 features}, each contributing one NT1-versus-control Cohen's $d$ and
one Benjamini--Hochberg FDR-corrected $q$ value
(Supplementary~\Cref{fig:to_nt1_supp}).

\begin{table}[!htbp]
\centering
\scriptsize
\caption{\textbf{Preprocessed feature library across cohorts.}
Feature families exposed through the shared data substrate, grouped by pipeline origin. Human-led denotes features or pipelines selected or provided by investigators; agent-led denotes pipelines curated, adapted, or implemented by the Preprocessing Agent under expert review and does not imply that the underlying method was invented by the agent. \checkmark\ indicates availability for at least a subset of eligible recordings; -- indicates that the feature is unavailable in the current substrate. Recording-level availability may vary because of missing signals and feature-specific quality control. SSC = Stanford Sleep Clinic; HSP = Human Sleep Project~\citep{li2026hsp}; CDH = Central Disorders of Hypersomnolence.}
\label{tab:features}
\resizebox{\textwidth}{!}{%
\begin{tabular}{llcccc}
\toprule
\textbf{Origin} & \textbf{Feature} & \textbf{SSC} & \textbf{HSP} & \textbf{BioSerenity} & \textbf{CDH} \\
\midrule
\multirow{10}{*}{Human-led}
 & Sleep stages (U-Sleep~\citep{perslev2021u})                              & \checkmark & \checkmark & \checkmark & \checkmark \\
 & Cortical arousals (MAD~\citep{brink2020automatic})                       & \checkmark & \checkmark & \checkmark & \checkmark \\
 & Respiratory events (ABED~\citep{kjaer2026expert})                        & \checkmark & \checkmark & \checkmark & \checkmark \\
 & EEG spectrograms                                                         & \checkmark & \checkmark & \checkmark & \checkmark \\
 & Transient oscillations (DynamO~\citep{stokes2023transient})              & --         & --         & --         & \checkmark \\
 & Brain--heart coupling~\citep{Saibene2026BrainHeart}                      & \checkmark & \checkmark & \checkmark & \checkmark \\
 & Demographics                                                             & \checkmark & \checkmark & \checkmark & \checkmark \\
 & Diagnoses (ICD or curated outcomes)                                      & \checkmark & \checkmark & --         & \checkmark \\
 & Medications                                                              & \checkmark & --         & --         & --         \\
 & Sleep questionnaires                                                     & --         & --         & \checkmark & --         \\
\midrule
\multirow{10}{*}{Agent-led}
 & R-peaks                                                                  & \checkmark & \checkmark & \checkmark & \checkmark \\
 & Spindles (YASA~\citep{vallat2021open})                                   & \checkmark & \checkmark & \checkmark & \checkmark \\
 & Slow waves (YASA~\citep{vallat2021open})                                 & \checkmark & \checkmark & \checkmark & \checkmark \\
 & Eye movements (YASA~\citep{vallat2021open})                              & \checkmark & \checkmark & \checkmark & \checkmark \\
 & EEG artefact masks (YASA~\citep{vallat2021open})                         & \checkmark & \checkmark & \checkmark & \checkmark \\
 & Continuous oximetry features                                             & \checkmark & \checkmark & \checkmark & \checkmark \\
 & Desaturation events                                                      & \checkmark & \checkmark & \checkmark & \checkmark \\
 & Heart-rate variability                                                   & \checkmark & \checkmark & \checkmark & --         \\
 & Brain--EMG coupling                                                      & \checkmark & \checkmark & --         & --         \\
 & Time-delay-stability networks (TDS~\citep{bashan2012network})            & \checkmark & \checkmark & --         & --         \\
\bottomrule
\end{tabular}}
\end{table}

\begin{table}[!htbp]
\centering
\footnotesize
\setlength{\tabcolsep}{4pt}
\caption{\textbf{Characteristics of the four PSG cohorts represented in the shared data substrate.}
$N$ denotes the number of PSG recordings in each cohort-level snapshot before case-study-specific eligibility, feature-availability, and quality-control criteria were applied.
Continuous variables are reported as mean $\pm$ SD among recordings with the corresponding variable available, and sex and race/ethnicity as percentages. Race/ethnicity follows cohort-native coding and may not sum to 100\% because of missing or unreported values. The analytical samples used in individual case studies are reported separately in the corresponding cohort-flow tables.}
\label{tab:cohort_demographics}
\begin{tabular}{p{3.7cm}rrrr}
\toprule
Variable & SSC & HSP & BioSerenity & CDH \\
\midrule
$N$ (PSG recordings) & 35{,}018 & 19{,}590 & 67{,}830 & 1{,}832 \\
Age, mean $\pm$ SD (years) & 45.5 $\pm$ 19.8 & 51.9 $\pm$ 16.6 & 51.2 $\pm$ 18.1 & 31.1 $\pm$ 15.9 \\
Age range (years) & 0--99 & 1--100 & 1--103 & 3--90 \\
Female sex (\%) & 40.6\% & 43.3\% & 48.5\% & 42.9\% \\
BMI, mean $\pm$ SD (kg/m$^{2}$) & 28.2 $\pm$ 7.2 & -- & 33.1 $\pm$ 9.1 & 25.3 $\pm$ 5.6 \\
\midrule
\multicolumn{5}{l}{\textit{Race/ethnicity (\%; cohort-native coding)}} \\
\quad White & 54.0\% & 77.6\% & 69.3\% & -- \\
\quad Asian & 12.9\% & 3.4\% & 0.6\% & -- \\
\quad African American & 2.9\% & 6.2\% & 23.1\% & -- \\
\quad Hispanic & 9.2\% & -- & 2.8\% & -- \\
\quad Other & 21.0\% & 10.2\% & 0.9\% & -- \\
\bottomrule
\end{tabular}
\end{table}

\begin{table}[!htbp]
    \centering
    \small
    \caption{\textbf{Cohort flow and disorder-group counts for SSC and HSP.}
    (a) Sequential exclusions used to derive the post-TDS-QC cohorts. All counts represent unique subjects.
    (b) PheCode-defined group counts before and after the common TDS quality filter. Cohort $n$ denotes subjects meeting the group definition after demographic eligibility but before TDS quality control; post-QC $n$ denotes subjects who additionally passed the TDS quality filter. Feature-specific analytic samples may be smaller because of missing physiological signals.
    $^{\dagger}$The Control group required the absence of documented neurological conditions (epilepsy, Parkinson's disease, dementia, Alzheimer's disease, and MCI); neuropsychiatric conditions (MDD and PTSD); cardiovascular conditions (hypertension, atrial fibrillation, heart failure, acute myocardial infarction, and ischaemic heart disease); sleep disorders (insomnia and REM sleep behaviour disorder); and stroke or TIA.
    $^{\ddagger}$Generalized convulsive epilepsy (PheCode 345.11) and convulsions (PheCode 345.3) were excluded from the epilepsy and partial-epilepsy groups. The generalized-convulsive-epilepsy exclusion affected 408 SSC and 232 HSP subjects.
    Disorder groups were evaluated independently and could overlap, except that subjects with both MDD and PTSD were excluded from the individual MDD and PTSD comparisons and tracked separately. Dashes indicate groups that were not analysed separately after TDS quality control. HSP did not include BMI; HSP models were therefore adjusted for age and sex only.}
    \label{tab:np_cohort_flow}

    \begin{tabular}{@{}clrrrr@{}}
        \toprule
        \multicolumn{6}{@{}l}{\textbf{a. Sequential cohort flow}} \\
        \addlinespace[2pt]
        & & \multicolumn{2}{c}{\textbf{SSC (discovery)}} &
            \multicolumn{2}{c}{\textbf{HSP (replication)}} \\
        \cmidrule(lr){3-4} \cmidrule(l){5-6}
        \textbf{Step} & \textbf{Criterion} &
        \textbf{Excluded} & \textbf{$n$ after} &
        \textbf{Excluded} & \textbf{$n$ after} \\
        \midrule
        0 & Starting subjects                         & --      & 39{,}920 & --  & 17{,}977 \\
        1 & Precomputed TDSpy output available        & 4{,}868 & 35{,}052 & 0   & 17{,}977 \\
        2 & Age and sex available                     & 14      & 35{,}038 & 0   & 17{,}977 \\
        3 & TDS feature-loading quality filter        & 1{,}836 & 33{,}202 & 167 & 17{,}810 \\
        \midrule
        \multicolumn{2}{@{}l}{\textbf{Final post-QC cohort}} &
        & \textbf{33{,}202} & & \textbf{17{,}810} \\

        \midrule
        \multicolumn{6}{@{}l}{\textbf{b. PheCode-defined group counts}} \\
        \addlinespace[2pt]
        & & \multicolumn{2}{c}{\textbf{SSC (discovery)}} &
            \multicolumn{2}{c}{\textbf{HSP (replication)}} \\
        \cmidrule(lr){3-4} \cmidrule(l){5-6}
        \multicolumn{2}{@{}l}{\textbf{Group}} &
        \textbf{Cohort $n$} & \textbf{Post-QC $n$} &
        \textbf{Cohort $n$} & \textbf{Post-QC $n$} \\
        \midrule
        \multicolumn{2}{@{}l}{Control$^{\dagger}$}                    & 11{,}809 & 11{,}240 & 3{,}632 & 3{,}604 \\
        \multicolumn{2}{@{}l}{Epilepsy$^{\ddagger}$}                  & 688       & 678       & 420     & 412 \\
        \multicolumn{2}{@{}l}{Partial epilepsy$^{\ddagger}$}          & 500       & 494       & 334     & 328 \\
        \multicolumn{2}{@{}l}{Parkinson's disease}                    & 629       & 612       & 303     & 302 \\
        \multicolumn{2}{@{}l}{Alzheimer's disease}                    & 361       & 352       & 432     & 429 \\
        \multicolumn{2}{@{}l}{Dementia}                               & 1{,}006   & 986       & 966     & 958 \\
        \multicolumn{2}{@{}l}{MCI}                                    & 1{,}242   & 1{,}207   & 1{,}030 & 1{,}019 \\
        \multicolumn{2}{@{}l}{PTSD}                                   & 280       & 268       & 244     & 242 \\
        \multicolumn{2}{@{}l}{MDD}                                    & 7{,}417   & 7{,}018   & 4{,}965 & 4{,}923 \\
        \multicolumn{2}{@{}l}{MDD+PTSD, excluded from both comparisons} & 921     & --        & 1{,}055 & -- \\
        \bottomrule
    \end{tabular}
\end{table}

\begin{table}[!htbp]
    \centering
    \small
    \caption{\textbf{Stage-1 quality cohort flow for SSC (development) and HSP v2.0 (external validation).}
    Sequential exclusion counts through the pre-specified QC pipeline: adult recordings ($\geq$ 18 y) with known demographics, at least one hypnodensity channel, recording duration $>$ 4 h, and total sleep time (TST) $>$ 90 min. TST is computed from the hypnodensity/hypnogram. Recordings that pass stage-1 QC are further excluded at the feature-extraction stage where modality-specific features are unavailable, so the analytic samples reported elsewhere are smaller than the split sizes tabulated here. The SSC test split contains 4{,}120 recordings, of which 4{,}096 carried the features required for the reported multidomain late-fusion evaluation (\Cref{fig:aging}a); the analytic \textit{n} for each individual domain clock is annotated in Supplementary~\Cref{fig:aging_supp}a.}
    \label{tab:aging_cohort_flow}
    \begin{tabular}{@{}lrrrr@{}}
        \toprule
        & \multicolumn{2}{c}{SSC} & \multicolumn{2}{c}{HSP v2.0} \\
        \cmidrule(lr){2-3} \cmidrule(l){4-5}
        QC step & \textit{n} excluded & \textit{n} remaining & \textit{n} excluded & \textit{n} remaining \\
        \midrule
        Starting cohort (canonical split)             & --      & 35{,}018            & --       & 19{,}590            \\
        Age known                                     & 3       & 35{,}015            & 1{,}613  & 17{,}977            \\
        Age $\geq$ 18                                 & 3{,}799 & 31{,}216            & 427      & 17{,}550            \\
        Sex coded                                     & 10      & 31{,}206            & 0        & 17{,}550            \\
        Hypnodensity present ($\geq$ 1 channel)       & 1{,}787 & 29{,}419            & 209      & 17{,}341            \\
        Recording duration $>$ 4 h                    & 322     & 29{,}097            & 258      & 17{,}083            \\
        Total sleep time $>$ 90 min                   & 280     & 28{,}817            & 110      & 16{,}973            \\
        \midrule
        \textbf{Stage-1 quality cohort (final)}       &         & \textbf{28{,}817}   &          & \textbf{16{,}973}   \\
        \midrule
        \multicolumn{5}{@{}l}{\textit{Canonical split assignments (subjects entering downstream modelling):}} \\
        \quad Train                                   &         & 24{,}098            &          & 8{,}932             \\
        \quad Validation                              &         & 599                 &          & 1{,}268             \\
        \quad Test                                    &         & 4{,}120             &          & 6{,}773             \\
        \bottomrule
    \end{tabular}
\end{table}

\begin{table*}[!htbp]
    \centering
    \caption{\textbf{Robustness of the sleep-ageing findings across the original analysis and three independent \method reruns.}
    To test whether the principal conclusions depended on a particular agent trajectory, the complete study was rerun three times from the same high-level prompt, with each run independently constructing its own feature space and analysis workflow. For the original run and each rerun, the table reports the quantities most relevant to the study's conclusions, together with an overall assessment. Values separated by ``/'' denote SSC and HSP, respectively. Feature counts are descriptive rather than directly comparable across runs, because each run independently engineered its feature space. Abbreviations: MAE, mean absolute error; HR, hazard ratio per 1\,SD residual; $R^2$, coefficient of determination; C-index, Harrell's concordance index; PheWAS, phenome-wide association study. The cardiometabolic composite comprised acute myocardial infarction, heart failure, and stroke/TIA. PheWAS replication was evaluated among SSC FDR-significant incident PheCodes tested in HSP, reported as Spearman $\rho$ and direction concordance. In the assessment column, ``replicated'' indicates the conclusion held across all runs, ``partially replicated'' the same qualitative pattern with quantitative variation, and ``negative result replicated'' that residual-only disease tuning failed to improve discrimination in every run. The Rerun~3 SSC composite HR was approximated from its forest plot.}
    \label{tab:aging_replication}
    \resizebox{\textwidth}{!}{%
    \begin{tabular}{@{}lccccc@{}}
        \toprule
        \textbf{Measure} & \textbf{Original} & \textbf{Rerun 1} & \textbf{Rerun 2} & \textbf{Rerun 3} & \textbf{Assessment} \\
        \midrule
        QC-passing recordings, SSC / HSP                  & 28{,}817 / 16{,}973 & 28{,}543 / 16{,}928 & 28{,}571 / 16{,}928 & 28{,}341 / 16{,}906 & Replicated \\
        Engineered features (count)                       & 362 & 622 & 1{,}105 & 365 & Variable \\
        Stacked-model age MAE, SSC / HSP (years)          & 7.06 / 8.83 & 8.00 / 13.20 & 6.79 / 8.51 & 8.41 / 11.46 & Partially replicated \\
        Stacked-model age $R^2$, SSC / HSP                & 0.686 / 0.494 & 0.55 / $-0.12$ & 0.71 / 0.53 & 0.56 / 0.10 & Partially replicated \\
        Cardiometabolic composite HR (per 1\,SD), SSC / HSP & 1.23 / 1.26 & 1.15 / 1.21 & 1.18 / 1.15 & $\approx$1.22 / 1.15 & Replicated \\
        Incident PheWAS replication, $\rho$ / concordance & 0.33 / 92.0\% & 0.60 / 95.0\% & 0.54 / 88.7\% & 0.55 / 88.4\% & Replicated \\
        Residual-only disease-tuned C-index, SSC / HSP    & 0.517 / 0.495 & 0.555 / 0.452 & 0.473 / 0.447 & 0.542 / 0.547 & Negative result replicated \\
        \bottomrule
    \end{tabular}%
    }
\end{table*}

\begin{table}[!htbp]
    \centering
    \small
    \caption{\textbf{Cohort flow for the COMISA case study.}
    The BioSerenity source contained 540{,}941 clinical and sleep-questionnaire records; the associated PSG cohort comprised 67{,}830 recordings. All tabulated counts refer to PSG recordings. Two analytic samples were derived independently from this cohort using the same four-group definitions and excluding recordings with AHI 5--14. The microstructure screen ($n_1$) required hypnodensity, arousal, spindle, and slow-wave features, whereas the arousal-focused analysis ($n_2$) required arousal, respiratory-event, and hypnodensity data. The first exclusion count in each branch is calculated relative to the 67{,}830-recording PSG cohort.}
    \label{tab:comisa_cohort}

    \begin{tabular}{@{}lrr@{}}
        \toprule
        \textbf{Criterion} & \textbf{Excluded} & \textbf{Remaining} \\
        \midrule
        BioSerenity PSG cohort & -- & 67{,}830 \\

        \midrule
        \multicolumn{3}{@{}l}{\textbf{Microstructure screen ($n_1$)}} \\
        Four-group eligibility and all four microstructure domains
            & 34{,}510 & 33{,}320 \\
        Non-empty feature files and $\geq10$\,min sleep
            & 1{,}482 & 31{,}838 \\
        Plausible covariates
            & 27 & \textbf{31{,}811} \\

        \midrule
        \multicolumn{3}{@{}l}{\textbf{Arousal-focused analysis ($n_2$)}} \\
        Four-group eligibility and required arousal, respiratory-event,
        and hypnodensity data
            & 31{,}282 & 36{,}548 \\
        Recording duration $\geq3$\,h
            & 155 & 36{,}393 \\
        At least 10 arousal events
            & 356 & 36{,}037 \\
        Total sleep time $\geq0.5$\,h
            & 73 & \textbf{35{,}964} \\
        \bottomrule
    \end{tabular}
\end{table}

\begin{figure}[p]
\centering
\small
\textbf{Step 1.}\; The script declares its inputs, outputs, and runtime in a machine-readable header.
\begin{lstlisting}[style=agentpy,firstnumber=18]
Inputs:
  - /workspace/execution/results/features_ssc.parquet   (28,817 × 421)
  - /workspace/execution/results/features_s0001.parquet (16,973 × 421)
  - /workspace/execution/results/feature_dictionary.csv (397 features)

Outputs:
  - /workspace/execution/models/clock_<c>.pkl  (8 clocks)
  - /workspace/execution/models/agg_global.pkl
  - /workspace/execution/models/agg_combined_meta.pkl
  - /workspace/execution/results/age_clock_performance.csv
  - /workspace/execution/results/age_clock_stratified_performance.csv
  - /workspace/execution/results/predictions_ssc.parquet
  - /workspace/execution/results/predictions_s0001.parquet
  - /workspace/execution/results/residuals_ssc.parquet
  - /workspace/execution/results/residuals_s0001.parquet
  - /workspace/execution/results/bias_correction_params.csv
  - /workspace/execution/results/residual_structure.csv
  - /workspace/execution/results/hrv_clip_caps.json
  - /workspace/execution/results/ahi_sensitivity_clocks.csv
  - /workspace/execution/results/stacker_bootstrap_stability.csv
  - /workspace/execution/results/common_subject_performance.csv
  - /workspace/execution/figures/fig03_age_clock_grid.pdf
  - /workspace/execution/figures/fig04_residual_vs_age_grid.pdf

Runtime: ~20-30 min.
\end{lstlisting}
\textbf{Step 2.}\; All performance metrics come from a single helper.
\begin{lstlisting}[style=agentpy,firstnumber=186]
def perf_metrics(y_true: np.ndarray, y_pred: np.ndarray) -> Dict[str, float]:
    mask = ~(np.isnan(y_true) | np.isnan(y_pred))
    y_true, y_pred = y_true[mask], y_pred[mask]
    if len(y_true) < 5:
        return dict(n=int(len(y_true)), mae=float("nan"), rmse=float("nan"),
                    r2=float("nan"), slope=float("nan"), intercept=float("nan"))
    mae = mean_absolute_error(y_true, y_pred)
    rmse = float(np.sqrt(mean_squared_error(y_true, y_pred)))
    r2 = r2_score(y_true, y_pred)
    slope, intercept = calibration_slope_intercept(y_true, y_pred)
    return dict(n=int(len(y_true)), mae=float(mae), rmse=rmse, r2=float(r2),
                slope=slope, intercept=intercept)
\end{lstlisting}
\textbf{Step 3.}\; Each row is labelled with the model that produced it.
\begin{lstlisting}[style=agentpy,firstnumber=514]
perf_rows = []
for cohort, y, pred in [
    ("ssc_train", y_train, p_train),
    ("ssc_val", y_val, p_val),
    ("ssc_test", y_test, p_test),
]:
    m = perf_metrics(y, pred)
    m.update(clock="combined_meta", cohort=cohort, estimator="ridge_stacker")
    perf_rows.append(m)
\end{lstlisting}
\textbf{Step 4.}\; The assembled table is written to the result file named in the header.
\begin{lstlisting}[style=agentpy,firstnumber=1402]
perf_df.to_csv(RESULTS_DIR / "age_clock_performance.csv", index=False)
\end{lstlisting}
\textbf{Step 5.}\; The reported value is read from that file, at full precision.
\begin{lstlisting}[style=agentout]
clock,cohort,estimator,n,mae,rmse,r2,...
combined_meta,ssc_test,ridge_stacker,4096,7.057098180235453,9.044013335769586,0.6860743374930721,...
\end{lstlisting}
\caption{\textbf{Code-grounding of a reported quantity: multidomain sleep ageing.}
The main text reports an MAE of 7.06 years and $R^2 = 0.686$ for the multidomain late-fusion model on the held-out SSC test set. The steps trace that quantity through \texttt{stage2\_iter7\_train\_clocks.py}: the declared inputs and outputs, the single helper that computes all performance metrics, the call that labels each row with the model producing it, the line that writes the result file, and the row from which the reported value is taken at full precision. Line numbers are those of the original script in the preserved workspace; excerpts are contiguous, indentation is reduced, and the result file is truncated at the quoted columns.}
\label{fig:code_grounding_aging}
\end{figure}

\begin{figure}[p]
\centering
\small
\textbf{Step 1.}\; The header declares inputs, outputs, and the parameters governing the analysis.
\begin{lstlisting}[style=agentpy,firstnumber=4]
Purpose: Cox proportional-hazards (incident cases) and logistic regression (all cases)
         to quantify the risk associated with lower TDS coupling for each comorbidity.
Inputs:
  - /workspace/execution/results/tds_features_ssc.parquet: per-subject TDS features (33,202 x 397)
  - /workspace/execution/results/cohort_ssc.csv: cohort with group assignments (35,038 rows)
  - /workspace/execution/results/cohort_summary.json: summary statistics
  - Expert-curated outcome data via get_SSC_stanford_outcome_df()
  - Phecode outcome data via get_SSC_stanford_phecode_outcome_df()
Outputs:
  - /workspace/execution/results/risk_analysis_hr.csv: all HR results
  - /workspace/execution/results/risk_analysis_or.csv: all OR results
  - /workspace/execution/results/risk_analysis.csv: combined summary
  - /workspace/execution/figures/fig06_hazard_ratios.pdf: forest plot of HRs
  - /workspace/execution/figures/fig07_odds_ratios.pdf: forest plot of ORs
Parameters:
  - penalizer: 0.01 -- regularization for CoxPHFitter to aid convergence
  - FDR q-threshold: 0.05 -- Benjamini-Hochberg correction within each comorbidity
  - BMI imputation: median -- for missing BMI values (13% of SSC)
  - Direction: predictors are NEGATED (per-SD decrease) so HR>1 = lower coupling -> higher risk
\end{lstlisting}
\textbf{Step 2.}\; A model is fitted for every predictor and sleep stage in turn.
\begin{lstlisting}[style=agentpy,firstnumber=213]
for pred_base, pred_display in PREDICTORS.items():
    for stage in STAGES:
        col = f"{pred_base}_{stage}"
        if col not in cox_data.columns:
            continue

        # Build Cox dataframe
        cox_df = cox_data[[col, 'age', 'sex', 'bmi', 'time_to_event', 'label']].dropna()
\end{lstlisting}
\textbf{Step 3.}\; The predictor is standardized and negated.
\begin{lstlisting}[style=agentpy,firstnumber=231]
# NEGATE: we want HR per SD DECREASE in coupling
# If we negate the z-scored predictor, HR > 1 means lower coupling -> higher risk
cox_df['tds_predictor'] = -((cox_df[col] - pred_mean) / pred_std)

cox_df_fit = cox_df[['tds_predictor', 'age', 'sex', 'bmi', 'time_to_event', 'label']].copy()
cox_df_fit.columns = ['tds_predictor', 'age', 'sex', 'bmi', 'T', 'E']

try:
    cph = CoxPHFitter(penalizer=PENALIZER)
    cph.fit(cox_df_fit, duration_col='T', event_col='E')

    hr = np.exp(cph.params_['tds_predictor'])
    ci = np.exp(cph.confidence_intervals_.loc['tds_predictor'].values)
\end{lstlisting}
\textbf{Step 4.}\; Each fitted model contributes one row, and the table is written to the result file.
\begin{lstlisting}[style=agentpy,firstnumber=258]
hr_results.append({
    'comorbidity': grp_name,
    'display_name': display,
    'predictor': pred_base,
    'predictor_display': pred_display,
    'stage': stage,
    'feature': col,
    'network': NETWORK_DISPLAY[pred_base],
    'HR': hr,
    'HR_lower': hr_lower,
    'HR_upper': hr_upper,
    'p_value': p_val,
    'n_total': len(cox_df_fit),
    'n_events': int(cox_df_fit['E'].sum()),
    'ph_assumption_ok': ph_ok,
    'pred_mean': pred_mean,
    'pred_std': pred_std,
})
\end{lstlisting}
\begin{lstlisting}[style=agentpy,firstnumber=399]
hr_df.to_csv(OUT_HR, index=False)
\end{lstlisting}
\textbf{Step 5.}\; The reported hazard ratio is read from that file, at full precision.
\begin{lstlisting}[style=agentout]
comorbidity,predictor,stage,HR,HR_lower,HR_upper
Parkinsons,mean_ls_brain_brain,N2,1.4778671263332086,1.3066306803136443,1.6715444356259153
\end{lstlisting}
\caption{\textbf{Code-grounding of a reported quantity: network-physiology coupling.}
The main text reports a hazard ratio of 1.48 (95\% confidence interval 1.31 to 1.67) for subsequent Parkinson's disease per 1\,SD decrease in N2 brain--brain coupling. The steps trace that quantity through \texttt{stage4\_risk\_analysis.py}. Line numbers are those of the original script; excerpts are contiguous, indentation is reduced, and the result file is shown for the quoted columns.}
\label{fig:code_grounding_network}
\end{figure}

\begin{figure}[!htbp]
    \centering
    \includegraphics[width=\linewidth]{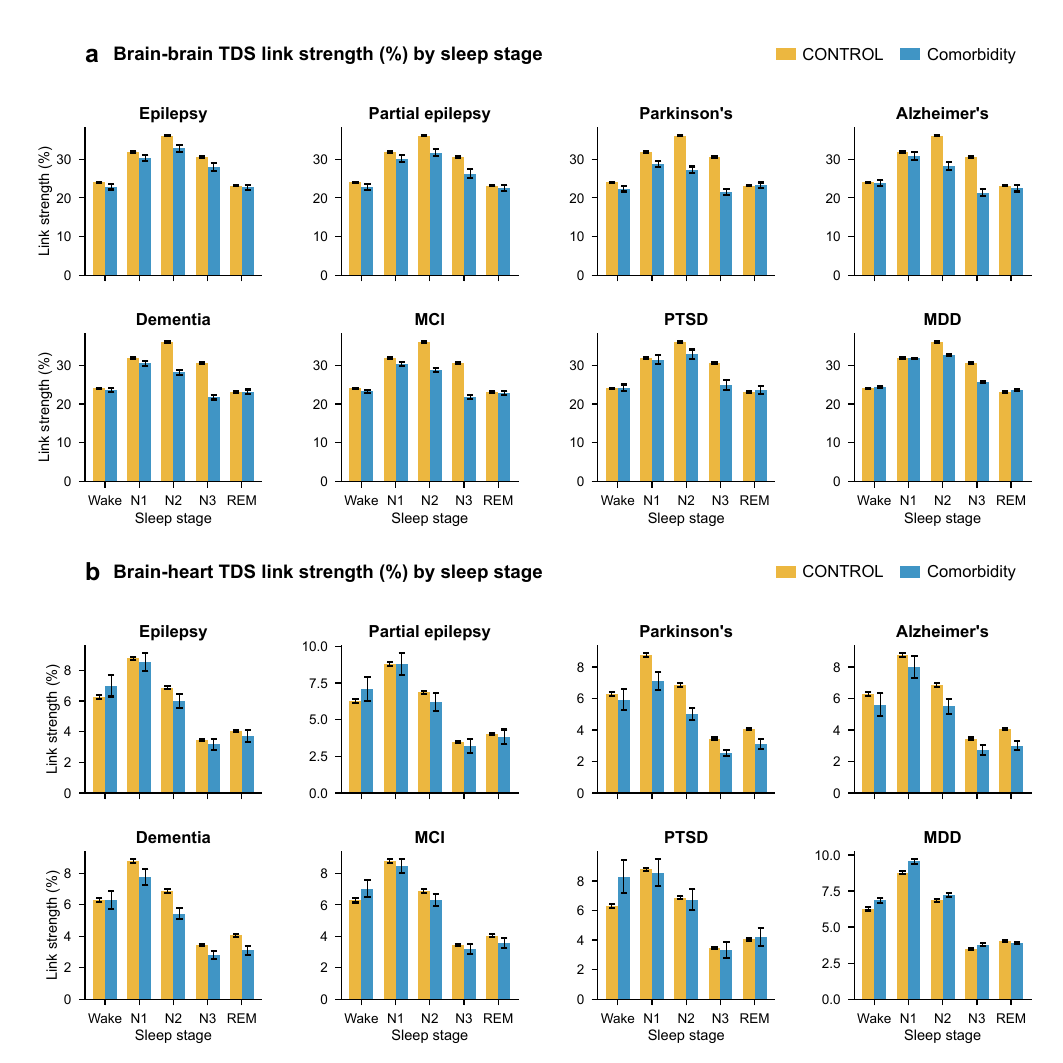}
    \caption{\textbf{Sleep-stage profiles of TDS link strength by disorder group in SSC.}
    (a) Brain--brain and (b) brain--heart TDS link strength (\%) across sleep stages (Wake, N1, N2, N3, and REM), shown for SSC controls and each of the eight disorder groups: epilepsy, partial epilepsy, Parkinson's disease, Alzheimer's disease, dementia, MCI, PTSD, and MDD. Bars show unadjusted group means, and error bars show 95\% confidence intervals. Each panel displays one disorder group alongside the SSC Control group; covariate-adjusted comparisons are presented in Supplementary~\Cref{fig:np_supp_volcano}.}
    \label{fig:np_supp_stage}
\end{figure}

\begin{figure}[!htbp]
    \centering
    \includegraphics[width=\linewidth,height=0.8\textheight,keepaspectratio]{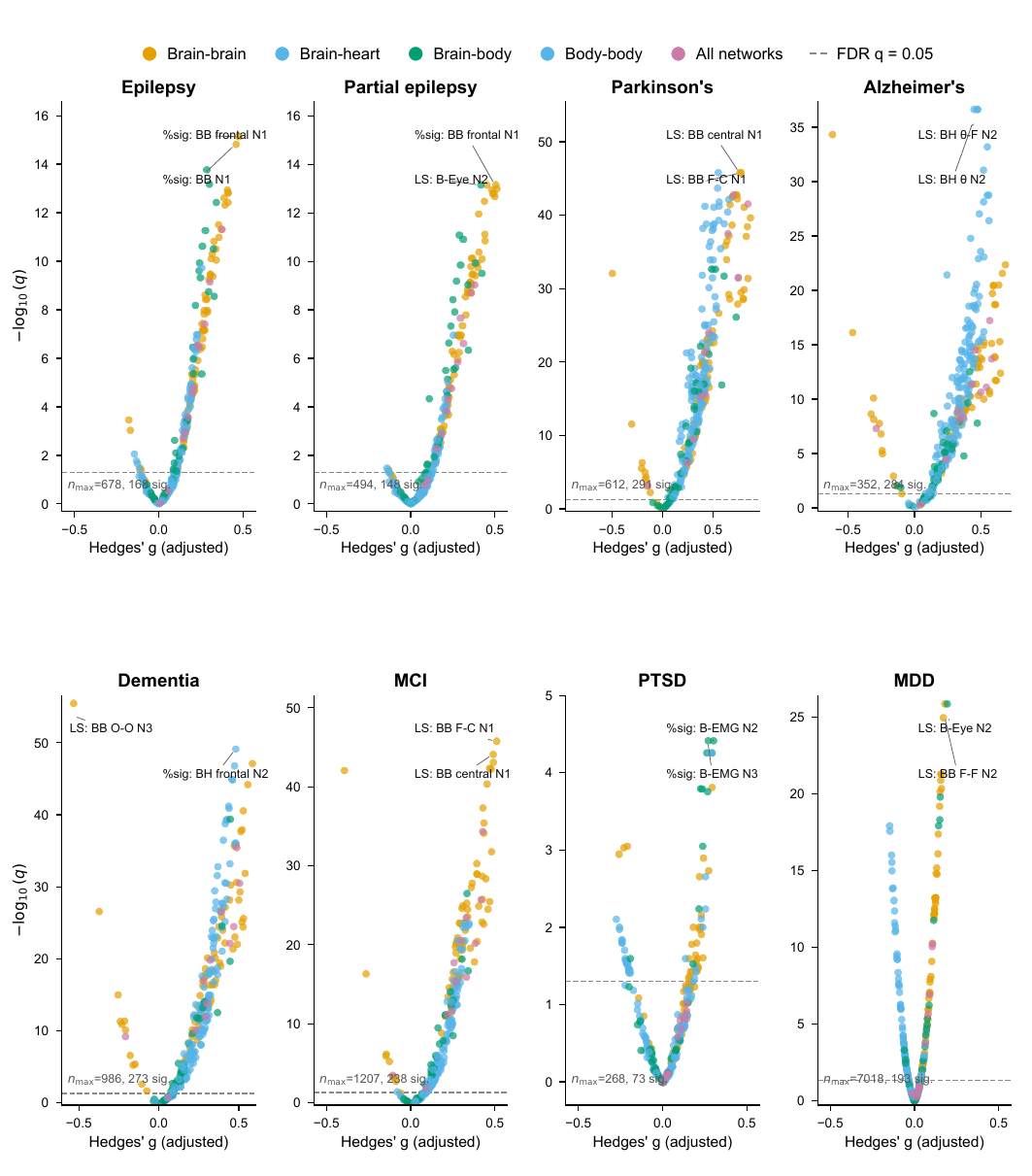}
    \caption{\textbf{Covariate-adjusted TDS coupling differences between disorder groups and controls in SSC.}
    Volcano plots show Hedges' $g$, calculated from feature residuals after adjustment for age, sex, and BMI, against FDR-adjusted statistical significance ($-\log_{10} q$) for each of the eight disorder groups. Positive $g$ indicates lower coupling in the disorder group than in controls, whereas negative $g$ indicates higher coupling. Each point represents a TDS coupling feature and is coloured by subnetwork (brain--brain, brain--heart, brain--body, body--body, or all-network summary). The dashed line marks the Benjamini--Hochberg FDR threshold ($q=0.05$). Each panel reports $n_{\max}$, the largest per-feature analytic sample for that group, and the number of FDR-significant features; the analytic sample varies across features according to feature and covariate availability. The two labelled features in each panel have the smallest FDR-adjusted $q$-values and are not necessarily those with the largest absolute effect sizes. LS, link strength; \%sig, percentage of significant links; BB, brain--brain; BH, brain--heart; B-EMG, brain--EMG; B-Eye, brain--eye; F, frontal; F-C, frontal--central.}
    \label{fig:np_supp_volcano}
\end{figure}

\begin{figure}[!htbp]
    \centering
    \includegraphics[width=\linewidth]{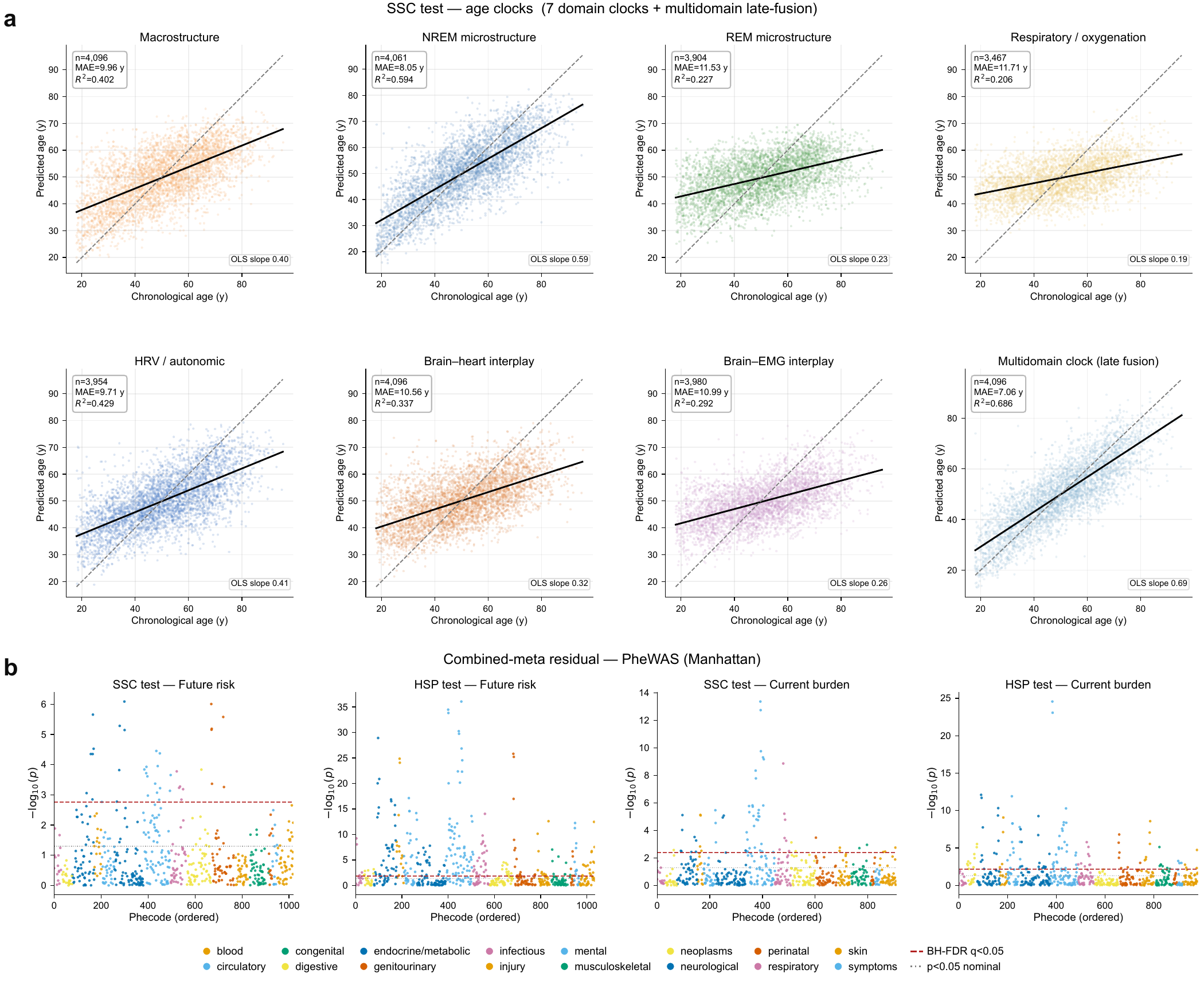}
    \caption{\textbf{Supplementary analyses for the multidomain sleep-ageing study.}
    (a) Age-prediction performance of the seven physiologically restricted domain clocks and the multidomain late-fusion clock on the held-out SSC test set. Each panel shows predicted versus chronological age, the identity line (grey dashed) and the fitted OLS calibration line (black solid), annotated with the analytic sample size, MAE, $R^2$, and calibration slope. Sample size varies across clocks because each requires its own input modalities. NREM microstructure was the strongest single-domain clock (MAE 8.05 years, $R^2=0.594$), and the multidomain late-fusion clock performed best overall (MAE 7.06 years, $R^2=0.686$). All calibration slopes were below 1, consistent with the age-dependent bias that the subsequent linear bias correction addresses (\Cref{fig:aging}a).
    (b) Phenome-wide association study (PheWAS) of the multidomain late-fusion age residual, labelled \emph{combined-meta} in the panel headers, shown as Manhattan plots for incident associations (future risk) and prevalent associations (current burden) in SSC and HSP. Each point is one PheCode, ordered along the $x$~axis and coloured by disease category; the red dashed line marks the Benjamini--Hochberg FDR threshold ($q<0.05$) and the grey dotted line nominal significance ($p<0.05$). Note that the $y$-axis range differs between panels. In SSC, 39 incident and 80 prevalent PheCodes reached FDR significance, concentrated in the circulatory, endocrine/metabolic, renal, genitourinary, and respiratory categories.}
    \label{fig:aging_supp}
\end{figure}

\begin{figure}[!htbp]
    \centering
    \includegraphics[width=\linewidth]{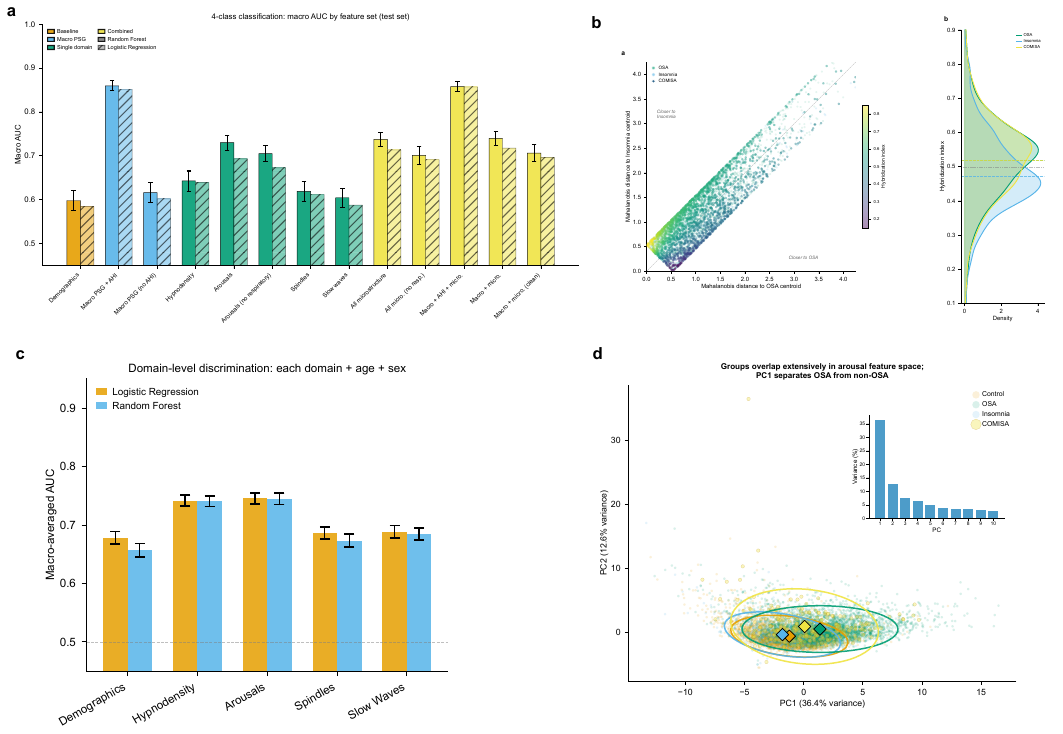}
    \caption{\textbf{Supplementary arousal-dynamics analyses for the COMISA replication cohorts.}
    (a) Four-class classification performance (macro-averaged AUC, test set) across feature sets ranging from demographics and macro-PSG baselines to single microstructure domains and combined feature sets, for random-forest and logistic-regression classifiers.
    (b) Position of the COMISA phenotype along the insomnia--OSA spectrum: (left) Mahalanobis distance of each recording to the OSA and insomnia centroids, coloured by hybrid score, and (right) distribution of the hybrid score for OSA, insomnia, and COMISA, showing the COMISA distribution shifted towards the OSA end.
    (c) Domain-level discrimination, shown as macro-averaged AUC for each microstructure domain combined with age and sex, for random-forest and logistic-regression classifiers.
    (d) Principal-component analysis of the arousal-dynamics feature space. The diagnostic groups overlap extensively, with PC1 (36.4\% of variance) separating OSA from non-OSA recordings. Inset, the variance explained by the first ten principal components.}
    \label{fig:comisa_supp}
\end{figure}

\begin{figure}[!htbp]
    \centering
    \includegraphics[width=0.85\linewidth, height=\textheight, keepaspectratio]{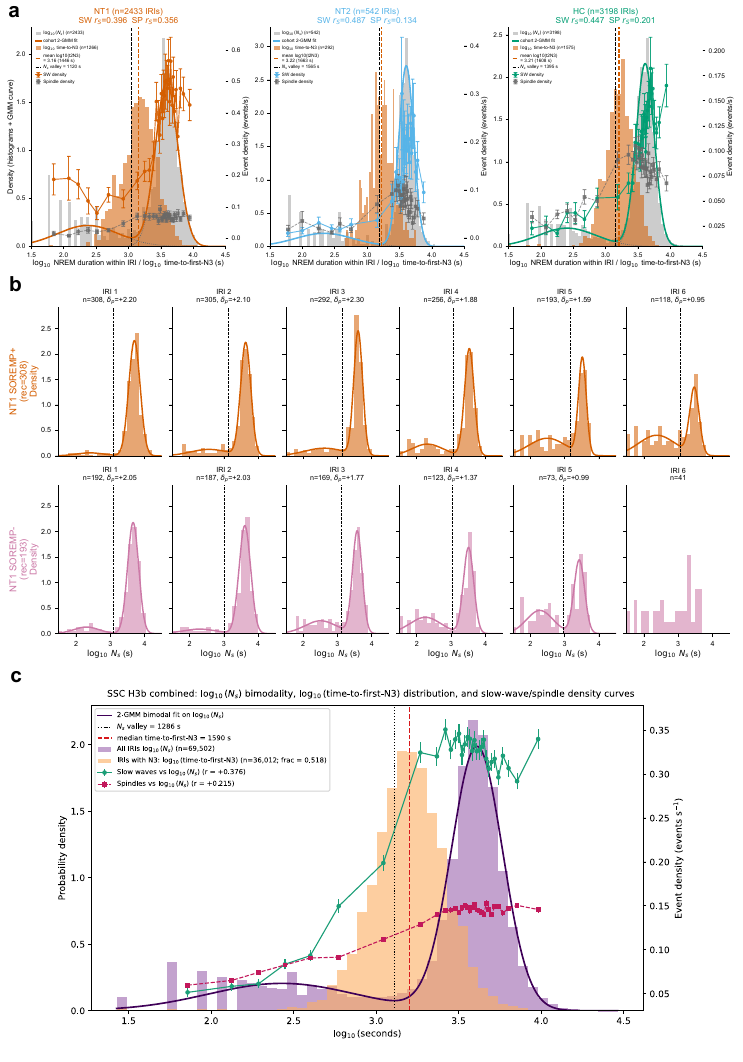}
  \caption{\textbf{Replication of homeostatic REM-pressure findings in CDH and SSC cohorts.}
  (a) Bimodality of $\log(N_s)$ and first-N3 alignment in CDH-NT1, CDH-NT2, and CDH controls.
  (b) Sleep-cycle-by-sleep-cycle evolution of the distribution of NREM duration during IRIs in NT1 patients, split by presence of a sleep-onset REM period (SOREMp) at the beginning of the night (first row, SOREMp+; second row, SOREMp$-$).
  (c) Bimodality of $\log(N_s)$ and first-N3 alignment in SSC.}
  \label{fig:rem_sleep_supp}
\end{figure}

\end{appendices}

\end{document}